\documentclass[twocolumn,showpacs,preprintnumbers,amsmath,amssymb,epsf,superscriptaddress]{revtex4-1}
\usepackage{graphicx,epsfig}% Include figure files
\usepackage{dcolumn}% Align table columns on decimal point
\usepackage{bm}% bold math
\usepackage{color}
\usepackage{comment}
\begin{document}

%\preprint{ICN/000-HEP}

\title{Deriving the slip front propagation velocity with the slip- and slip-velocity-dependent friction laws via the use of the linear marginal stability hypothesis}
\author{Takehito Suzuki}
\email{t-suzuki@phys.aoyama.ac.jp}
%\author{Hiroshi Matsukawa}
\affiliation{Department of Physical Sciences, Aoyama Gakuin University, 5-10-1 Fuchinobe, Chuo-ku, Sagamihara, Kanagawa 252-5258, Japan}

%\date{\today} % It is always \today, today,

\begin{abstract}
We analytically and numerically investigate the determining factors of the slip front propagation (SFP) velocity. The slip front has two forms characterized by intruding or extruding front. We assume a 1D viscoelastic medium on a rigid and fixed substrate, and we employ the friction law depending on the slip and slip velocity. Despite this dependency potentially being nonlinear, we use the linear marginal stability hypothesis, which linearizes the governing equation for the slip, to investigate the intruding and extruding front velocities. The analytically obtained velocities are found to be consistent with the numerical computation where we assume the friction law nonlinearly depends on both the slip and slip velocity. This implies that the linearized friction law is sufficient to capture the dominant features of SFP behavior.%We found two intruding SFP velocities, and three extruding SFP velocities. Three of these five velocities describe the optical modes, and the single intruding SFP velocity and the single extrusing SFP velocity are physically meaningful. 
\end{abstract}

%%%%%\pacs{98.70.Sa, 98.38.Mz, 96.50.sb}% PACS, the Physics and Astronomy                     
\maketitle

\section{Introduction} \label{secInt}

%When we shear solid objects on the solid substrate and increase the shear force adiabatically, the macroscopic slip of the block occurs only when the shear force exceeds the critical value: the macroscopic static friction force. Though this behavior has been recognized for a long time, the mechanism determining the critical value has not been clarified yet. Moreover, local slip prior to the macroscopic sliding (referred to as precursors in the present paper) occurs in some systems. For example, many experimental (e.g., \cite{Ben}) and numerical (e.g., \cite{Ots}) studies show that local slip different from the macroscopic sliding appears. This slip has no stress singularity ahead of the front, so that it can be arrested by local small perturbation such as small increase of normal stress. From physical and industrial viewpoints, the unified framework must be established to understand systematically these complex behaviors about macroscopic slip initiation, even though it has not been achieved.

The slip front propagation (SFP) is an important phenomenon in several scientific and technological fields, including frictional sliding on the interface between two media and crack tip propagation in a medium. From a theoretical viewpoint, researchers have investigated SFP using both discrete \cite{Lan93, Amu1, Amu2, Tro} and continuum \cite{Rad, Suz19} models. Furthermore, laboratory experiments have also been performed to investigate SFP \cite{Ben}. Geophysical studies have also contributed to the understanding of the SFP behavior because fault tip propagation can be modeled as SFP. The SFP velocity has attracted considerable research interest. For example, ordinary and slow earthquakes are widely known to exist \cite{Oba, Kat}, and they have different SFP velocities; the SFP velocities for the former category of the earthquake are much higher than those of the latter. The origin of this difference remains an important open research question.%For example, SFP emerges in the system with numbers of elastic blocks connected by springs on the rigid substrate \cite{Lan93}. On the other hand, the continuum model has been treated to investigate the slip-front-propagation velocity \cite{Rad}. 

The friction law plays an important role in the understanding of SFP. For example, a friction stress depends on the state variable in the rate-and-state dependent friction model \cite{Amu2, Bar13, Rad}. Bar-Sinai \textit{et al.} \cite{Bar13} investigated the SFP velocity using a semi-infinite continuum model and the rate-and-state friction law. Furthermore, we note that the friction stress can also depend on quantities such as the slip and slip velocity. For example, Myers and Langer \cite{Mye} employed a spring-block model with a slip-velocity-weakening law to study the SFP velocity. The rate-and-state dependent friction law with the shear stiffness of the contacts can be interpreted as the slip-strengthening behavior \cite{Bar12}. Moreover, the friction law depending on the slip and slip velocity is realized for elasto- and visco-dampers \cite{Hib03, Hib05}. The slip- and slip-velocity- dependences of friction stress is important in geophysical studies \cite{Ika, Ito}.

Additionally, the viscosity of the medium is important in understanding the SFP behavior \cite{Ots, Mye, Suz19}. Our previous study \cite{Suz19} analyzed an infinitely long viscoelastic block on a rigid substrate and achieved the systematic understanding of the SFP velocity in terms of the viscoelasticity and the friction law depending on the slip velocity in a quadratic manner. Despite these studies, a thorough investigation of the SFP velocity with viscoelastic medium and a friction law that depends on the slip and slip velocity has not yet been undertaken. %

Spontaneous SFP velocity has been widely understood by regarding the phenomenon of SFP as the evolution of a solution of the governing equation from a stable state into an unstable state and employing the linear marginal stability hypothesis (LMSH) \cite{Lan91, Mye, Suz19}. The LMSH has been widely used to investigate the dynamics of fronts or domain walls that propagate spontaneously into an unstable state in a model with a nonlinear governing equation \cite{Saa1, Saa2}. This approach has been used to obtain the solidification front speed \cite{Dee}, the chemical reaction front speed \cite{Saa1, Saa2}, and the slip front velocity between blocks and substrates \cite{Lan91, Suz19, Mye}. This hypothesis asserts that even if the governing equations are nonlinear, a linearized model yields sufficiently accurate front behavior, including the correct front propagating velocity. This hypothesis has been applied to the slip front behavior via the linearization of the friction law; the friction laws adopted in such studies have been limited, e.g., the slip-velocity-weakening \cite{Lan91}; the systematic treatment of various types of friction law has not been performed.

%Front propagation of a domain has been attracted in many scientific fields such as solidification, chemical reaction, and slip between blocks and substrates. For example, KPP equation has been treated widely. Pattern formation has also been a controversial problem.

This paper is organized as follows. A brief introduction to the LMSH is provided in Sec.~\ref{secLMSH}, and the model setup is defined in Sec.~\ref{secMS}. Analytical treatment of the problem is presented in Sec.~\ref{secAT}; in this section several SFP velocities are obtained based on the LMSH, and a discussion of their physical implications are presented. The numerical treatment and implications for slip behavior are discussed in Sec.~\ref{secNT}, and the treatment of the LMSH is justified there. A summary and discussion are presented in Sec.~\ref{secDisCon}.

\section{MODEL WITH VELOCITY-DEPENDENT LOCAL FRICTION LAW} \label{secMWVD}

\subsection{Linear Marginal Stability Hypothesis (LMSH)}  \label{secLMSH}

The LMSH has been widely used to investigate the dynamics of fronts or domain walls that propagate spontaneously into an unstable state where the phenomena are described by nonlinear governing equations \cite{Saa1, Saa2}. This approach has been used to obtain the solidification front speed \cite{Dee}, the chemical reaction front speed \cite{Saa1, Saa2}, and the SFP velocity \cite{Mye, Suz19, Lan91}. We briefly summarize this approach here. The LMSH requires linearizing the governing equation, the plane wave approximation of the solution near the propagating front, and the two conditions associated with the stability of growth of the disturbance and that of propagation. This hypothesis states that the characteristic frequency, the wave number, and the propagating velocity of a front can be derived even after these approximations.

\begin{figure}[tbp]
\centering
\includegraphics[width=8.5cm]{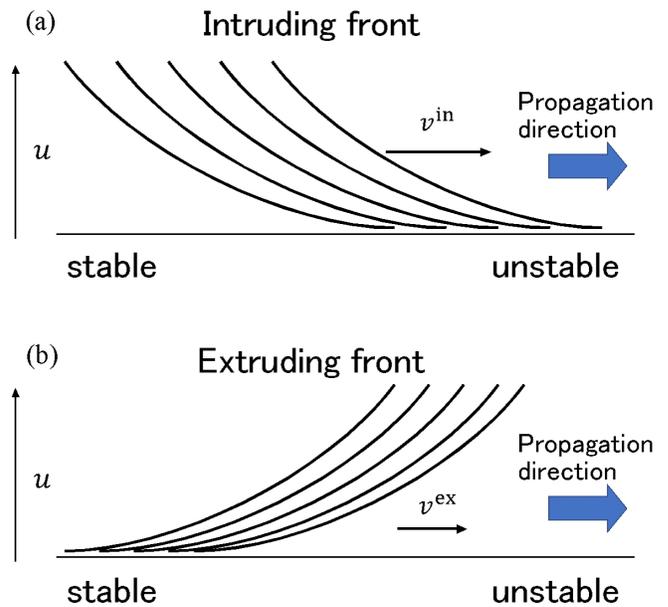}
\caption{Schematic illustrations of the intruding and extruding fronts.}
\label{FigEIF}
\end{figure}

To understand the details of LMSH further, we define $u$ as a variable characterizing the state of the system (in the case of motion of a continuum, this variable could be slip); we consider the dynamics of the spontaneous propagation of $u$. Here we assume an infinite and 1D system. We consider two cases: (1) $u=0$ is unstable and the stable region with $u>0$ intrudes into this unstable region, and (2), $u=0$ is stable and this region intrudes into the unstable region with $u>0$. The front in the former case is referred to as an ``intruding front,'' whereas that in the latter case is called an ``extruding front'' (Fig.~\ref{FigEIF}). Notably, if the governing equation of $u$ is a wave equation, the cases are symmetric and the propagation velocity is equal in magnitude. However, if the governing equation includes additional terms, such as a diffusion term, the symmetry vanishes, and different propagation velocities are expected for the two fronts.

The front is mathematically defined to be located where terms of $O(|u|)$ dominate and terms of $O(|u|^2)$ become negligible in the governing equations. For the solution of $u$, we assume the plane wave solution, $u \sim \exp(\pm i(kx-\omega t))$, whose frequency $\omega$ and wave number $k$ are complex in the region close to the front. The upper (lower) sign corresponds to the intruding (extruding) front propagation. To understand this propagation, we rewrite this plane wave solution of $u$ as
\begin{eqnarray}
\exp (\pm i (kx-\omega t))=\exp (\pm i ((k_r+i k_i)x-(\omega_r + i \omega_i)t)) \nonumber \\ 
= \exp (\pm i (k_r x - \omega_r t)) \exp (\mp (k_i x - \omega_i t)), \label{eqFF}
\end{eqnarray}
where $k_r$ and $k_i$ are the real and imaginary parts of $k$, respectively, and $\omega_r$ and $\omega_i$ are the real and imaginary parts of $\omega$, respectively. Notably, the absolute value of Eq.~(\ref{eqFF}) can determine the front, i.e., $\exp (\mp (k_i x - \omega_i t))$ is used to define the front. The value of $\exp (\pm i (k_r x - \omega_r t))$ describes the oscillation within the envelope $\exp (\mp (k_i x - \omega_i t))$, and is not used to determine the position of the front. Equation (\ref{eqFF}) and Fig.~\ref{FigEIF} clearly indicate that the upper (lower) sign corresponds to the intruding (extruding) front. In the above expression, $k_r, \ k_i, \ \omega_r$ and $\omega_i$ are four unknown parameters, and the LMSH determines them via four independent equations: the real and imaginary parts of the dispersion relation and the expressions for growth and propagating stabilities. With these values, the primary goal of this study is obtaining an analytical form for the intruding and extruding front velocity. 

%This description yields $|s|=\exp [\pm (k_i x - \omega_i t)]$, where $k_i$ and $\omega_i$ are the imaginary parts of $k$ and $\omega$, respectively, and are assumed to be nonnegative. Note that the fronts $\exp(k_i x -\omega_i t)$ and $\exp[-(k_i x -\omega_i t)]$ describe the extruding and intruding fronts, respectively, based on their definition. We have four unknown parameters: $k_i$, $\omega_i$, $k_r$ (the real part of $k$), and $\omega_r$ (the real part of $\omega$). The parameters $k_r$ and $\omega_r$ are also assumed to be nonnegative.

The growth stability condition mentioned above is given by
\begin{equation}
\frac{\partial \omega_i}{\partial k_r}=0,  \label{eqGI1}
\end{equation}
whereas, the propagating stability is given by the relationship:
\begin{equation}
\frac{\partial \omega_i}{\partial k_i}=\frac{\omega_i}{k_i}=c, \label{eqPI1}
\end{equation}
where $c$ is a positive constant. The remaining equations are given by
\begin{equation}
\frac{\partial \omega_r}{\partial k_i} =0, \label{eqGI2}
\end{equation}
and
\begin{equation}
\frac{\partial \omega_r}{\partial k_r} =\frac{\omega_r}{k_r} = c, \label{eqPI2}
\end{equation}
which are obtained from the Cauchy-Riemann relationship.

Equation (\ref{eqGI1}) represents the condition that the disturbance will not grow with increasing time. Equation (\ref{eqPI1}) indicates that the phase velocity $c_p \equiv \omega_i/k_i$ is equal to the group velocity $c_g \equiv \partial \omega_i/\partial k_i$. Given that the disturbance propagates with the group velocity, this condition dictates that the disturbance and the front propagate with the same velocity. It is important to note that for stability, the relationship $c_p \ge c_g$ is sufficient because the disturbance is overtaken by the front with $c_p \ge c_g$. Nonetheless, Ref.~\cite{Saa1} mathematically showed that Eq.~(\ref{eqPI1}) is satisfied for spontaneous front propagation. For more details regarding these conditions see Ref.~\cite{Suz19}. 

\subsection{Model setup} \label{secMS}

%We consider the continuous limit of the uniform Burridge-Knopoff model including infinite numbers of blocks connected by springs on a substrate. All the blocks are connected with the upper substrate by springs. The model has been employed by several studies to investigate SFP velocities \cite{Mye, Lan91}. We use this model with the slip- and slip-velocity-dependent friction laws. We apply the loading stress on the left end in the direction tangential to the substrate surface, i.e., end-loading stress. The end-loading stress is written as $p_e(<0)$, and the stress is applied along the $x$ axis for the time $t \ge 0$. Therefore, the system is considered to be 1D along the $x$ axis, and the slip distance $u(x, t)$ at position $x$ and time $t$ is zero throughout the whole system for $t <0$. With this setup, we can consider the nondimensionalized governing equation

In this work we consider a semi-infinite viscoelastic medium that is assumed to be on the rigid and fixed substrate. We apply a loading stress on the left end of the medium in the direction tangential to the substrate surface, i.e., an end-loading stress. We employ the nondimensionalization of the variables based on Ref.~\cite{Suz19}. The end-loading stress is denoted $p_e(<0)$, and this constant stress is applied along the $x$ axis for a time $t \ge 0$. We also consider additional loading conditions. Among the loading conditions, we evaluate in this work the continuum limit of the Burridge-Knopoff model, which includes a driving stress from the upper block connected to the upper plate \cite{Mye}. We refer to this loading stress as ``upper-loading stress.'' The upper-loading stress is denoted $p_u (>0)$, and this constant stress is applied along the $x$ axis for the time $t > -\infty$. We assume that $p_u$ is less than the macroscopic maximum static friction stress, which is assumed to be the same for all the blocks considered. We treat the two models (with the different loading conditions) in the single framework: the model subject to the end-loading stress is referred to as the EL model, whereas the model with end- and upper-loading stresses is called the EUL model. In these models, the system is considered to be 1D along the $x$ axis, and the slip distance, $u(x, t)$, at position $x$ and time $t$ is zero throughout the whole system for $t <0$. 

We can consider the nondimensionalized governing equation:% We treat the semi-infinite viscoelastic block on a rigid substrate, and The condition realizing this model setup is described in Ref. \cite{Suz19} in detail. Actually, the boundary conditions can be selected arbitrary \cite{Mye}. The discretized and continuum models can generate SFP. We should treat such SFPs in the single framework.
\begin{equation}
\ddot{u}=u''+\dot{u}''+F(u)-\tau_{\mathrm{v}}(\dot{u})-\tau_{\mathrm{s}}(u), \label{eqNUFG}
\end{equation}
where $F(u)$ is a slip-dependent function describing the loading condition, $\tau_{\mathrm{v}}(\dot{u})$ and $\tau_{\mathrm{s}}(u)$ describe the local friction stresses that depend on $\dot{u}$ and $u$, respectively, and the dot and prime represent differentiation with respect to $t$ and $x$, respectively. The loading conditions can be treated systematically via the function $F(u)$. We can select $F(u)=0$ for the EL model \cite{Suz19}, and $F(u)=p_u-u$ for the EUL model \cite{Mye}. The terms $\ddot{u}$, $u''$, and $\dot{u}''$ have a physical interpretation; they represent the inertia, elastic, and viscous terms, respectively. This viscous term is employed in several previous studies \cite{Mye, Suz19}. %We require $\tau_{\mathrm{v}}(0)+\tau_{\mathrm{s}}(0)=1$, and the value unity correponds to the static friction stress. Additionally, the explicit loading parameter $\sigma$ on the right hand side of Eq. (\ref{eqNUFG}) corresponds to the shear stress acting the blocks when $\dot{u}=0$ and $u=0$ \cite{Mye}. Therefore, $\sigma=1$ is the critical case, where infinitely small disturbance induces the slip propagation.

%Actually, this model setup is also the expansion of the model of Ref. \cite{Suz19}, which assumed the infinitely long viscoelastic block on the rigid substrate. The equation of motion (\ref{eqNUFG}) and the boundary condition can also describe the slip profile for such a system, as shown below.

Several forms of $\tau_{\mathrm{v}}(\dot{u})$ and $\tau_{\mathrm{s}}(u)$ have been considered in the analysis of SFP (e.g., \cite{Lan91, Mye, Suz19}), and the SFP velocity has been analytically investigated previously \cite{Suz19}. The purpose of the present paper is to derive a more universal expression for SFP velocity, which is independent of the details of friction laws; such an analysis enables us to understand the SFP behavior in the single framework.

\section{Analytical treatment} \label{secAT}

We now linearize the terms $F(u)-\tau_{\mathrm{v}}(\dot{u})-\tau_{\mathrm{s}}(u)$ from Eq.~(\ref{eqNUFG}). We take
\begin{equation}
F(u)-\tau_{\mathrm{v}}(\dot{u})-\tau_{\mathrm{s}}(u)=C_1 \dot{u} -C_2 u, \label{eqLin}
%F(u)-\tau_{\mathrm{v}}(\dot{u})-\tau_{\mathrm{s}}(u)=C_{\mathrm{v}}^{\mathrm{fric}} \dot{u} -C_{\mathrm{s}}^{\mathrm{fric}} u,
\end{equation}
where $C_1$ and $C_2$ are constants. 
%where $C_{\mathrm{v}}^{\mathrm{fric}}$ and $C_{\mathrm{s}}^{\mathrm{fric}}$ are constant numbers. 
To undertake the linearization (\ref{eqLin}), we assume that the system is in a critical state before the end-loading in the EUL model, which leads to
\begin{equation}
p_u=\tau_{\mathrm{v}}(0)+\tau_{\mathrm{s}}(0). \label{eqCr}
\end{equation}
Analyses without the critical state assumption have been performed in previous studies \cite{Mye, Bar12}. However, those treatments only provide approximations for the SFP velocities. With the assumption of the critical state, an analytical treatment is possible here.%In addition, we do not consider the case where the slip direction reverses ($u$ can turns to be negative) with increasing time, as done in seismology \cite{Suz06}.

We note that if the slip direction reverses ($u$ turns to be negative) with increasing time, Eq.~(\ref{eqLin}) cannot be applied. The reason for this is twofold: firstly, the friction stress, $\tau_{\mathrm{v}}(\dot{u})+\tau_{\mathrm{s}}(u)$, changes its sign if slip reversal occurs, and thus, the linearization of Eq.~(\ref{eqLin}) cannot be employed. Secondly, if the slip reversal occurs, the absolute value of the stress acting on the slip-reversal point will exceed the maximum static friction stress. This slip criterion is not considered in the linearization of Eq.~(\ref{eqLin}). We do not consider the slip reversal in the following analytical treatment, as is done in seismology \cite{Suz06}. %Notably, the reversal of the slip direction does not occur for the EL model.%This corresponds to the case where $F(u)=\sigma -u$ turns to be negative. This sign change of the friction stress will be called ``stress-sign-change.''  We call this criterion ``slip criterion,'' and and we will employ the stress-sign-change and the slip criterion in the numerical calculations.We will find that the friction laws exemplified in the numerical calculations do not induce the slip reversal, and judge that the analytical treatment completely works. Treating the slip reversal will be a potential future work. 

%Actually, for the EUL model, the upper-loading stress can act in the negative $x$ direction, which may induce the reversal of the slip direction. If the slip direction reverses, Eq. (\ref{eqLin}) cannot be applied. However, we should consider that $F(u)=-\sigma-u$ must be negative to induce the slip reversal. This corresponds to the condition that the static friction stress must be exceeded when the slip reversal occurs. Since we assume the critical state (\ref{eqCr}) implying that the upper-loading stress acting in the positive $x$ direction is equal to the static friction stress, we assume that the slip reversal does not occur, and the slip velocity is assumed to be always positive. Notably, the reversal of the slip direction does not occur for the EL model. %This statement will be confirmed by numerical calculations in Sec. \ref{secNT} for several friction laws. 
%Notably, the state is not critical to the $-x$ direction. 

%We also define $C_1 \equiv C_{\mathrm{v}}^{\mathrm{fric}}$ and $C_2 \equiv 1+C_{\mathrm{s}}^{\mathrm{fric}}$ in the following discussion. 
Using the linearization of Eq.~(\ref{eqLin}), we rewrite the governing equation [Eq.~(\ref{eqNUFG})],
\begin{equation}
\ddot{u} = u''+ \dot{u}'' +C_1 \dot{u} -C_2 u.  \label{eqeom4}
\end{equation}
%If $C_1$ and $C_2$ are positive, comparing the results with those of previous studies \cite{Lan93, Suz19} is easy with this form.
We note positive and negative values for $C_1$ and $C_2$ are permitted. 

We assume the plane wave solution for Eq.~(\ref{eqeom4}), i.e., $u \sim \exp(\pm i(kx-\omega t))$. The upper (lower) sign describes the intruding (extruding) front propagation, as noted in Sec.~\ref{secLMSH}. Based on this assumption and using Eq.~(\ref{eqeom4}), the dispersion relation is found to be%If $C_2>0$, the system constructed by \cite{Lan93} is realized. 
\begin{equation}
- \omega^2 = -k^2 \pm i \omega k^2 \mp i C_1 \omega -C_2.  \label{eqdis2}
\end{equation}
Considering the real and imaginary parts of the dispersion relation in Eq.~($\ref{eqdis2}$), we obtain,
\begin{equation}
(\mp \omega_i - 1)(k_r^2 - k_i^2) \mp 2 \omega_r k_r k_i + (\omega_r^2 - \omega_i^2) \pm C_1 \omega_i -C_2 = 0, \label{eqD12}
\end{equation}
\begin{equation}
(\mp \omega_i - 1) \cdot 2 k_r k_i \pm \omega_r (k_r^2 - k_i^2) + 2 \omega_r \omega_i \mp C_1 \omega_r = 0, \label{eqD22}
\end{equation}
respectively. We first note that the solutions $k_r \neq 0$ and $\omega_r \neq 0$ exist. In the case of $C_1$ and $C_2$ being positive and $k_i=\omega_i=0$, the solution set $(k_r,\omega_r,k_i,\omega_i)=(\sqrt{C_1},\sqrt{C_1+C_2},0,0)$ exists. The solutions $k_r \neq 0$ and $\omega_r \neq 0$ describe an oscillating solution. The slip reversal occurs with these solutions; such solutions cannot be treated in a framework based on Eq.~(\ref{eqeom4}). Thus, this solution set is excluded from the following discussion. We therefore assume $k_r=\omega_r=0$ in the analytical treatment. With such an assumption, considering Eq.~($\ref{eqD12}$), we have
\begin{equation}
k_i^2 \pm \omega_i k_i^2 -C_2 \pm C_1 \omega_i - \omega_i^2 =0. \label{eqD3}
\end{equation}
Combining this equation, the expression for growth stability [Eq.~(\ref{eqGI1})] and condition for propagation stability [Eq.~(\ref{eqPI1})], we obtain a cubic equation for $\omega_i$,
\begin{equation}
\omega_i^3 \mp 2C_1 \omega_i^2 - (C_1-3C_2) \omega_i \pm 2C_2 = 0, \label{eqo3}
\end{equation}
and a relationship between $k_i$ and $\omega_i$:
\begin{equation}
k_i^2 = \frac{\mp 2C_2+C_1 \omega_i}{\omega_i}. \label{eqk2}
\end{equation}

We begin by categorizing the solutions of Eq.~(\ref{eqo3}) in terms of the numbers of positive and negative real solutions. Combining the results with Eq.~(\ref{eqk2}), we are able to investigate the numbers of SFP velocities that the above equations yield.

We start by defining $f_{\mathrm{in}} (\omega_i)$ as being equal to the left-hand side of Eq.~(\ref{eqo3}) with the upper sign; such a definition permits us to study the intruding front velocity. It is noted that the discussion performed in this section is also directly applicable to the extruding front propagation. To confirm this statement, we define $f_{\mathrm{ex}} (\omega_i)$ as the left-hand side of Eq.~(\ref{eqo3}) with the lower sign. We then have the relationship
\begin{eqnarray}
f_{\mathrm{in}} (-\omega_i) &=& -\omega_i^3 -2C_1 \omega_i^2 + (C_1-3C_2) \omega_i + 2C_2 \nonumber \\
&=& -f_{\mathrm{ex}} (\omega_i).
\end{eqnarray}
Thus, we see that the number of the real solutions for $f_{\mathrm{ex}} (\omega_i)=0$ is exactly the same as the number of real solutions for $f_{\mathrm{in}} (\omega_i)=0$, and the positive (negative) solutions for the former correspond to the negative (positive) solutions for the latter. We therefore investigate only the numbers of real solutions for $f_{\mathrm{in}} (\omega_i)$ in this work.

To categorize the numbers of the real solutions for $f_{\mathrm{in}} (\omega_i)=0$ in terms of $C_1$ and $C_2$, we require an expression for $D$, $f_{\mathrm{in}}(0)$, $f_{\mathrm{in}}(\omega_{\pm}^{\mathrm{in}})$ and $\omega_{\pm}^{\mathrm{in}}$, where
\begin{equation}
D \equiv 4C_1^2 + 3 (C_1-3C_2),  \label{beq5}
\end{equation}
and $\omega_{\pm}^{\mathrm{in}}$ is defined as the solution of
\begin{equation}
\frac{\partial}{\partial \omega_i} f_{\mathrm{in}} (\omega_i) \Big|_{\omega_i=\omega^{\mathrm{in}}_{\pm}} = 3 {\omega^{\mathrm{in}}_{\pm}}^2 - 4C_1 \omega^{\mathrm{in}}_{\pm} - (C_1-3C_2) = 0,  \label{beq2}
\end{equation}
which leads to the solution
\begin{equation}
\omega^{\mathrm{in}}_{\pm} = \frac{2C_1 \pm \sqrt{ 4 C_1^2 + 3 (C_1-3C_2) }}{3} \label{beq3}
\end{equation}
(double signs in the same order). We observe that if $D$ is positive $\omega^{\mathrm{in}}_{\pm}$ is real. Additionally, we can obtain
\begin{equation}
f_{\mathrm{in}}(0) = 2C_2 \label{beq1}
\end{equation}
and
\begin{equation}
f_{\mathrm{in}} (\omega^{\mathrm{in}}_{\pm}) = {\omega^{\mathrm{in}}_{\pm}}^3 - 2C_1 {\omega^{\mathrm{in}}_{\pm}}^2 -(C_1-3C_2) \omega^{\mathrm{in}}_{\pm} + 2C_2.  \label{beq4}
\end{equation}

\begin{figure}[tbp]
\centering
\includegraphics[width=8.5cm]{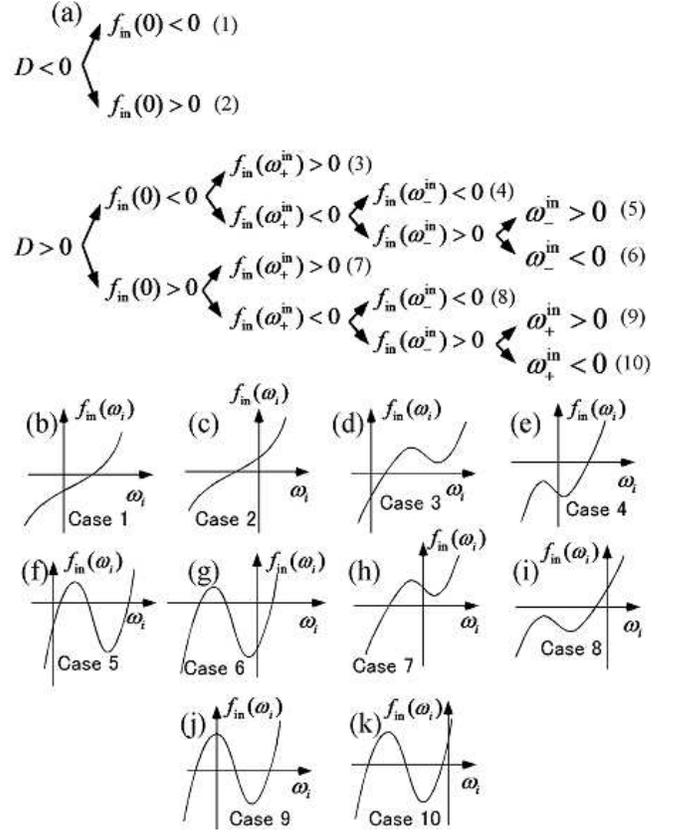}
\caption{(a) Depicting the various cases examined in this work. (b-k) Various forms of $f_{\mathrm{in}}(\omega_i)$ for the cases listed in (a).} %In (e) and (h), the solid curves show $f_{\mathrm{in}}(\omega_i)$ for $\omega_{\pm}^{\mathrm{in}}<0$, whereas the dashed curves show the condition $\omega_{\pm}^{\mathrm{in}}>0$. In (g) the solid curve shows $f_{\mathrm{in}}(\omega_i)$ for $\omega_+^{\mathrm{in}}<0$, whereas the dashed curve depicts $f_{\mathrm{in}}(\omega_i)$ for $\omega_+^{\mathrm{in}}>0$. In (j) the solid curve shows $f_{\mathrm{in}}(\omega_i)$ for $\omega_-^{\mathrm{in}}<0$, whereas the dashed curve shows $f_{\mathrm{in}}(\omega_i)$ for $\omega_-^{\mathrm{in}}>0$. The numbers of positive and negative solutions for $f_{\mathrm{in}}(\omega_i)=0$ are the same for the solid and dashed curves in (e), (g), (h), and (j).
\label{FigDia}
\end{figure}

We define $\omega_1^{\mathrm{in}}$, $\omega_2^{\mathrm{in}}$, and $\omega_3^{\mathrm{in}}$ as the solutions for $f_{\mathrm{in}}(\omega_i)=0$; $\omega_1^{\mathrm{in}}$ is defined to always take real values, as described in Appendix A. We number the various cases based on the diagram shown in Fig.~\ref{FigDia}(a). In cases 1, 2, 3, 4, 7, and 8, only $\omega_1^{\mathrm{in}}$ is real, and $\omega_1^{\mathrm{in}}$ is positive in cases 1, 3, and 4, whereas it is negative in cases 2, 7, and 8. Though the numbers of the positive (negative) solutions are the same for the cases 1, 3, and 4 (2, 7, and 8), we differentiate the cases because the numbers of the real positive solutions for $k_i$, which is one or zero, may be different in each case. The solutions $\omega_1^{\mathrm{in}}$, $\omega_2^{\mathrm{in}}$, and $\omega_3^{\mathrm{in}}$ are real in cases 5, 6, 9, and 10. All the solutions are positive in case 5, two are positive and one is negative in case 9, two are negative and one is positive in case 6, and all the solutions are negative in case 10. %Finally, we note that the sign of $\omega_+^{\mathrm{in}}$ in cases 4, 6, and 7 and that of $\omega_-^{\mathrm{in}}$ in cases 4, 7, and 9 do not affect the discussion below (see the dashed lines in Fig.~\ref{FigDia}(e), (g), (h), and (j)).

%The diagram based on $D$, $f_{\mathrm{in}}(0)$, $f_{\mathrm{in}}(\omega_{\pm}^{\mathrm{in}})$, and $\omega_{\pm}^{\mathrm{in}}$ shows that the behaviors of the solution of Eq. (\ref{eqo3}) are categorized into 10 cases (Fig. \ref{FigDia}). We can briefly summarize the numbers of real solutions for all cases. Cases 1, 3, and 4 have the single positive solution for $f_{\mathrm{in}} (\omega_i)=0$. Cases 2, 7 and 8 have the single negative solution. Case 5 has three positive solutions. Case 6 has two negative solutions and one positive solution. Case 9 has two positive solutions and one negative solution. Case 10 has three negative solutions. If the curve $f_{\mathrm{in}}(\omega_i)$ touches the $\omega_i$ axis, the case gives a boundary. 

We now obtain the analytical forms of the boundaries dividing cases 1--10 in $C_1-C_2$ phase space. These equations are obtained based on the diagram Fig.~$\ref{FigDia}$(a) as follows:
\begin{equation}
D=0,  \label{b0eq1}
\end{equation}
\begin{equation}
f_{\mathrm{in}} (0) =0,  \label{b0eq2}
\end{equation}
\begin{equation}
f_{\mathrm{in}} (\omega_+^{\mathrm{in}})=0,  \label{b0eq3}
\end{equation}
\begin{equation}
f_{\mathrm{in}} (\omega_-^{\mathrm{in}})=0,  \label{b0eq4}
\end{equation}
\begin{equation}
\omega_+^{\mathrm{in}}=0,  \label{b0eq5}
\end{equation}
\begin{equation}
\omega_-^{\mathrm{in}}=0.  \label{b0eq6}
\end{equation}
First, using Eqs.~($\ref{beq5}$) and ($\ref{b0eq1}$), we obtain a  boundary referred to as boundary A, which reduces to the parabolic form
\begin{equation}
C_2=\frac{4}{9} \left(C_1 + \frac{3}{8} \right)^2 -\frac{1}{16}.  \label{beqA}
\end{equation}
From Fig.~\ref{FigDia}, boundary A represents a boundary between a region in which solutions are characterized as cases 1--2 and a region containing solutions that fall into cases 3--10 in the $C_1-C_2$ space. Next, from Eqs.~(\ref{beq1}) and (\ref{b0eq2}), we can conclude that $C_2=0$ represents a boundary, which we will call boundary B. Boundary B divides regions of solutions that are represented as case 1 from case 2, and the solutions in cases 3--6 from cases 7--10. Further, we consider the cubic equation for $C_2$ from Eqs.~(\ref{beq4}), (\ref{b0eq3}), and (\ref{b0eq4}):
\begin{eqnarray}
3C_2^3 &-& (C_1^2 - 3C_1 -3) C_2^2 \nonumber \\
&-& (C_1^2 + \frac{10}{9} C_1^3) C_2-\frac{C_1^3}{9} -\frac{C_1^4}{9} =0.
\end{eqnarray}
We can factorize this equation to obtain,
\begin{equation}
(C_2+C_1 +1)(3C_2^2 - C_1^2 C_2 -\frac{C_1^3}{9})=0,
\end{equation}
which we can evaluate to find that
\begin{equation}
C_2=-C_1 - 1, \ \ \ \frac{C_1^2}{6} \pm \frac{1}{3} \sqrt{ \frac{C_1^4}{4} + \frac{C_1^3}{3}}
\end{equation}
are boundaries. The former will be referred to as boundary C, and the latter will be referred to as boundary $\mathrm{D}_{\pm}$, respectively (double signs in the same order). Boundary C is a boundary for regions where solutions are of case 3, case 4, and cases 5--6, and $\mathrm{D}_{\pm}$ divides the regions in which solutions in case 7, case 8, and cases 9--10 exist. Additionally, we can confirm analytically that the boundaries A, C, and $\mathrm{D}_+$ all pass through the point $(C_1, C_2)=(-1.5, 0.5)$, and that the boundary C is tangential to the boundaries A and $\mathrm{D}_+$ at this point.

\begin{figure}[tbp]
\centering
\includegraphics[width=8.5cm]{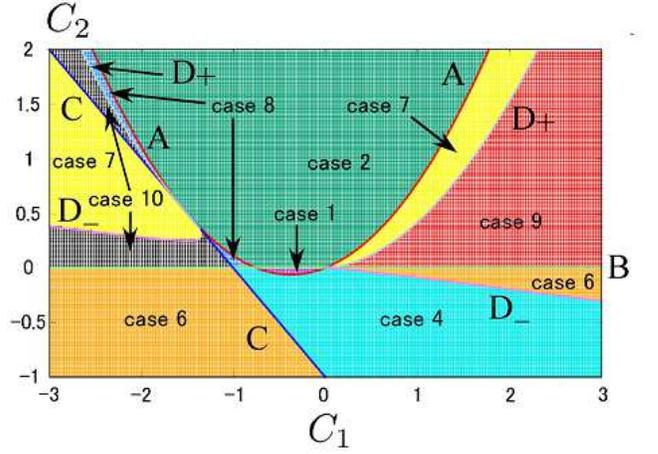}
\caption{Phase space for the solutions for $f_{\mathrm{in}}(\omega_i)=0$.}
\label{FigPha1}
\end{figure}

Using the boundaries $\mathrm{A}-\mathrm{D}_{\pm}$, we obtain the phase space that categorizes the behavior of the solutions for $f_{\mathrm{in}} (\omega_i)=0$ (see Fig.~\ref{FigPha1}). Figure \ref{FigPha1} shows that cases 6, 7, 8, and 10 are further divided into two disconnected regions. Actualy, from Eqs.~(\ref{beq3}), ($\ref{b0eq5}$), and ($\ref{b0eq6}$), we can obtain the relationship $C_2=C_1/ 3$ for $C_1>0$ [Eq.~($\ref{b0eq6}$)] and for $C_1<0$ [Eq.~(\ref{b0eq5})]. This forms a boundary between regions in which solutions for $f_{\mathrm{in}} (\omega_i)=0$ fall into cases 5 and 6, and cases 9 and 10. However, we can see that the case 5 does not appear in the phase space (see Fig.~\ref{FigPha1}), and case 9 is not adjacent to case 10. Therefore, this straight line does not appear as a boundary in Fig.~\ref{FigPha1}.  %The symbol ``-'' represents that the cases related by this symbol cannot be divided by this boundary. This straight line is the tangent of the parabolic line ($\ref{beqA}$) at the origin. 

\begin{figure}[tbp]
\centering
\includegraphics[width=9.5cm]{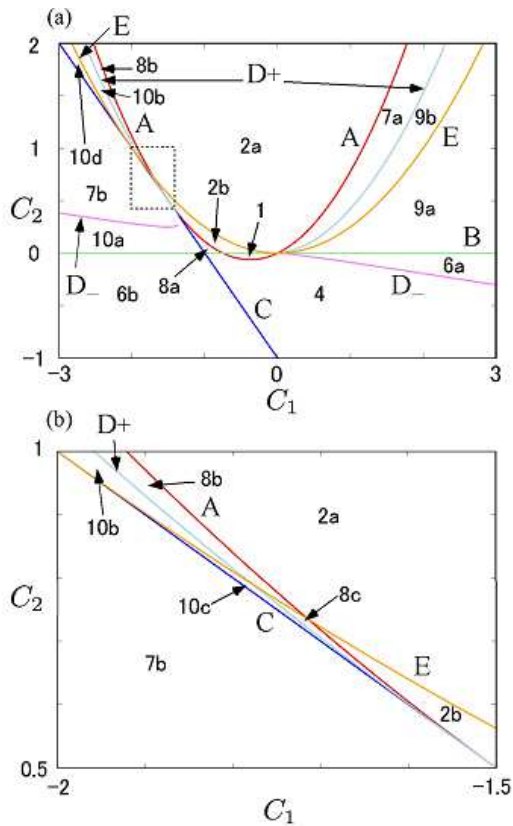}
\caption{Phase space with the boundaries A$-$E. The numbers and alphabets refer to the case of the solution. The region enclosed by a dotted line in (a) is enlarged in (b).}
\label{FigPha2}
\end{figure}

The solutions for $f_{\mathrm{in}}(\omega_i)=0$ have been categorized in the $C_1-C_2$ space. We now categorize the solutions for $k_i$. Notably, the solution for $k_i$ is obtained from Eq.~(\ref{eqk2}); thus we see that the solution does not exist when the right-hand side of Eq.~(\ref{eqk2}) is negative, even if real and positive solutions for $\omega_i$ exist. Therefore, we see that the curve $k_i=0$ is another boundary in the $C_1-C_2$ space. We refer to this boundary as boundary E, and the analytical form of the boundary E is obtained here. If we substitute $k_i=0$ to Eq.~(\ref{eqD3}) with the upper sign, we obtain %In actuality, to investigate the behaviors of such solutions, we need the other boundary. 
\begin{equation}
\omega_i= \frac{C_1}{2} \pm \sqrt{\frac{C_1^2}{4} -C_2},
\end{equation}
where the sign of the second term on the right-hand side is arbitrary. Substituting $k_i=0$ in Eq.~($\ref{eqk2}$) with the upper sign yields
\begin{equation}
\omega_i= \frac{2C_2}{C_1}.
\end{equation}
From the two previous expressions, an equation describing boundary E can be obtained:
\begin{equation}
C_2=\frac{C_1^2}{4}. \label{eqBE}
\end{equation}
It can also be confirmed that boundary C is a tangent to boundary E at the point $(C_1, C_2)=(-2, 1)$. The phase space is shown in Fig.~\ref{FigPha2}. This figure shows 17 subdivided regions of the cases, and they are named as shown in the figure (see Table \ref{Tcase} for details). %which is used in the subsequrent text.
%Note that the case 9b has the solution $v_3^{\mathrm{in}}$, whereas the case 9a does not have such a solution. Additionally, the cases 2a, 8b, and 10b have the solution $v_1^{\mathrm{ex}}$, whereas the cases 2b, 8c, and 10c do not have such a solution. Moreover, the case 10d shows the solution $v_3^{\mathrm{ex}}$, while the case 10b does not include the solution. on the boundary between the cases 2a and 2b, between the cases 9a and 9b, and between the cases 10b and 10d, which gives the analytical form of the boundary E. , indicating that the $C_1-C_2$ space is divided into 17 regions

We show the values of $\omega_j^{\mathrm{in}}$, $\omega_j^{\mathrm{ex}}$, $k_j^{\mathrm{in}}$, and $k_j^{\mathrm{ex}}$ ($j=1,2,3$) in Fig.~\ref{FigO-K}, where $\omega_j^{\mathrm{ex}}$, $k_j^{\mathrm{in}}$, and $k_j^{\mathrm{ex}}$ are defined in Appendix A. Note that $\omega_2^{\mathrm{in}}$ and $k_2^{\mathrm{in}}$ do not exist. The values of $k_3^{\mathrm{in}}$, $k_1^{\mathrm{ex}}$, and $k_3^{\mathrm{ex}}$ are zero on boundary E. Note that since Eq.~(\ref{eqBE}) is a necessary and not a sufficient condition for $k_j^{\mathrm{in}}=0$ or $k_j^{\mathrm{ex}}=0$, non-zero values ($k_1^{\mathrm{in}}$ and $k_2^{\mathrm{ex}}$) are allowed on bounary E. %$k_j^{\mathrm{in}}$ satisfies $(k_j^{\mathrm{in}})^2=(-2C_2+C_1 \omega_j^{\mathrm{in}})/\omega_j^{\mathrm{in}}$. If $k_3^{\mathrm{in}}$ is zero, $k_1^{\mathrm{in}}$ is not zero because $\omega_1^{\mathrm{in}} \neq \omega_3^{\mathrm{in}}$. Therefore, we obtain non-zero $k_1^{\mathrm{in}}$ on boudary E. %In terms of these solutions, the $C_1-C_2$ space is divided into 17 regions (Fig. \ref{FigTwo} and ). The solutions $k_j^{\mathrm{in}}$ and $k_j^{\mathrm{in}}$ are obtained from $\omega_j^{\mathrm{in}}$ and $\omega_j^{\mathrm{ex}}$ and Eq. (\ref{eqk2}).

\begin{figure*}[tbp]
\centering
\includegraphics[width=16cm]{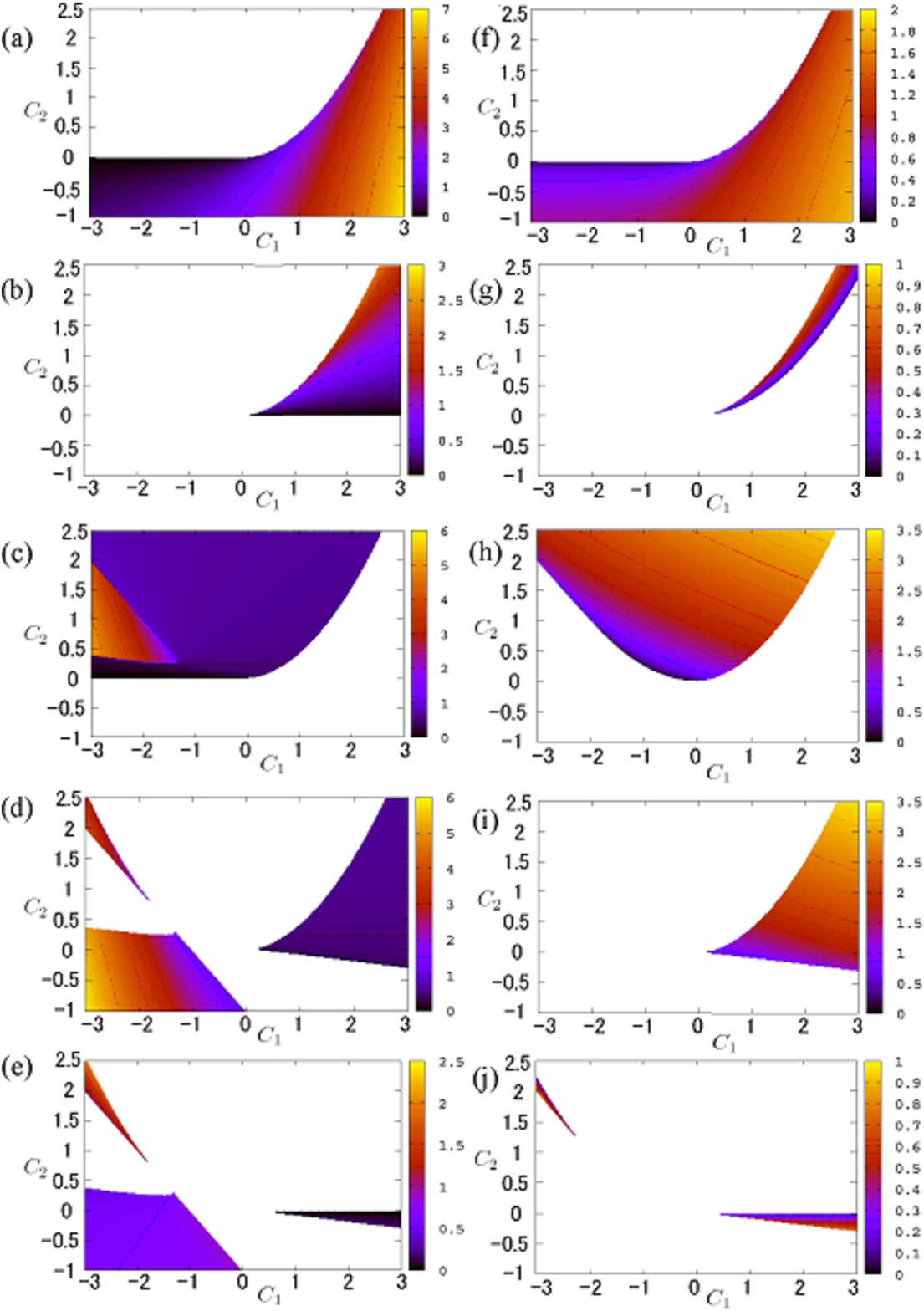}
\caption{Values for (a) $\omega_1^{\mathrm{in}}$, (b) $\omega_3^{\mathrm{in}}$, (c) $\omega_1^{\mathrm{ex}}$, (d) $\omega_2^{\mathrm{ex}}$, (e) $\omega_3^{\mathrm{ex}}$, (f) $k_1^{\mathrm{in}}$, (g) $k_3^{\mathrm{in}}$, (h) $k_1^{\mathrm{ex}}$, (i) $k_2^{\mathrm{ex}}$, (j) $k_3^{\mathrm{ex}}$. }%The bright area in (c) represents case 7b; the values in this area are discontinuous from the adjacent areas.}
\label{FigO-K}
\end{figure*}

\begin{figure*}[tbp]
\centering
\includegraphics[width=12.5cm]{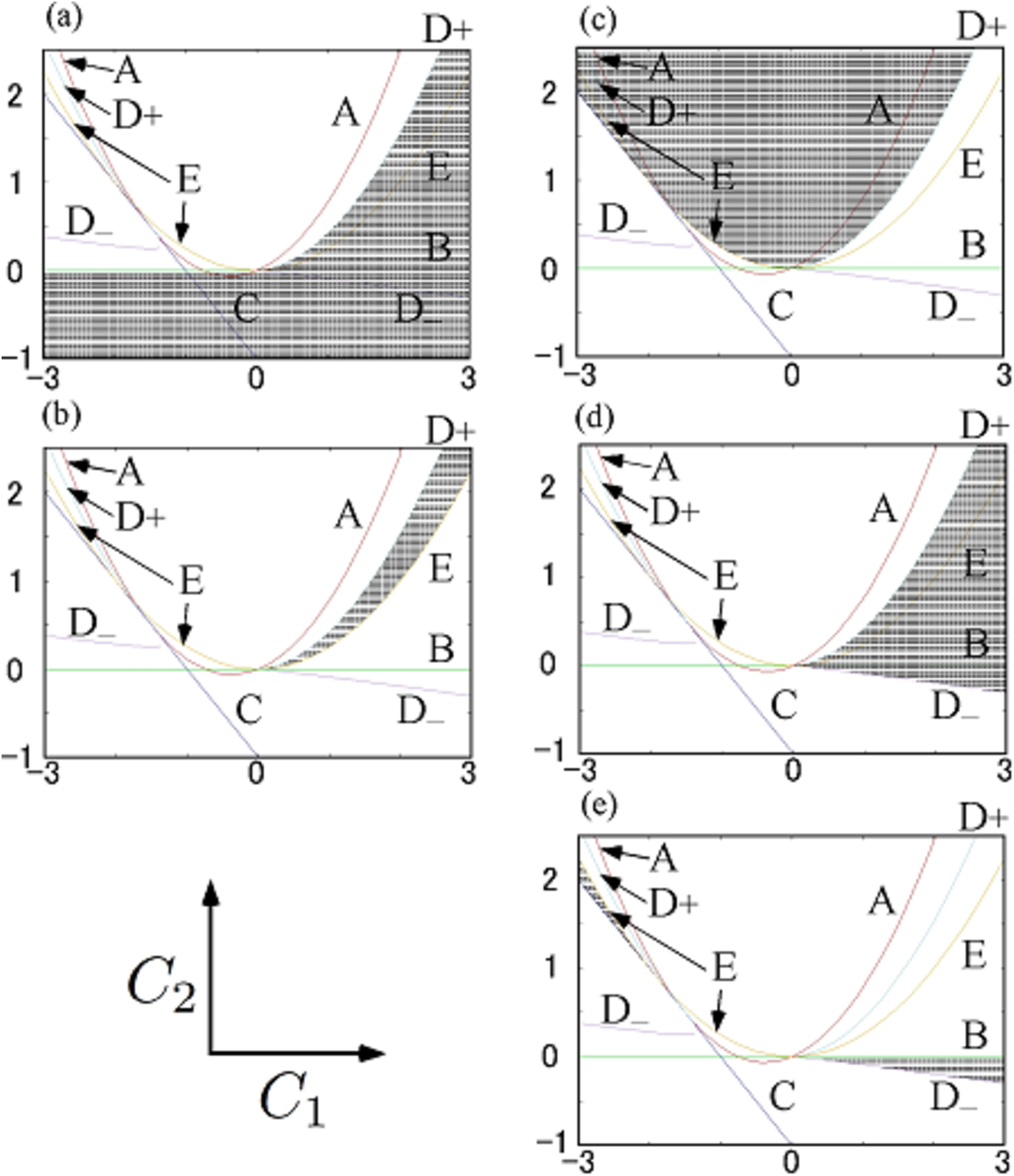}
\caption{The parameter regions for the front velocities. The shaded areas indicate the regions in which (a) $v_1^{\mathrm{in}}$, (b) $v_3^{\mathrm{in}}$, (c) $v_1^{\mathrm{ex}}$, (d) $v_2^{\mathrm{ex}}$, and (e) $v_{31}^{\mathrm{ex}}$ and $v_{32}^{\mathrm{ex}}$ are found to exist.}
\label{FigTwo}
\end{figure*}

Using the values of $\omega_j^{\mathrm{in}}$, $\omega_j^{\mathrm{ex}}$, $k_j^{\mathrm{in}}$, and $k_j^{\mathrm{in}}$ corresponding to solutions of $f_{\mathrm{in}}(\omega_i)=0$, we also investigate the intruding front velocity, $v^{\mathrm{in}}$, and the extruding front velocity, $v^{\mathrm{ex}}$. We define the $j$th intruding and extruding front velocity as $v_j^{\mathrm{in}}$ and $v_j^{\mathrm{in}}$, respectively, as in Appendix A, and the velocities $v^{\mathrm{in}}_1$, $v^{\mathrm{in}}_3$ $v^{\mathrm{ex}}_1$, $v^{\mathrm{ex}}_2$, and $v^{\mathrm{ex}}_3$ exist. The regions where $v_1^{\mathrm{in}}$, $v_3^{\mathrm{in}}$, $v_1^{\mathrm{ex}}$, $v_2^{\mathrm{ex}}$, and $v_3^{\mathrm{ex}}$ exist are shown in Fig.~\ref{FigTwo}(a--e), respectively, in $C_1-C_2$ space. We note that the region where $v_3^{\mathrm{ex}}$ exists is separated into two disconnected regions. We refer to $v_3^{\mathrm{ex}}$ which corresponds to values of negative $C_1$ as $v_{31}^{\mathrm{ex}}$, and that which corresponds to positive values of $C_1$ as $v_{32}^{\mathrm{ex}}$. We have three velocities in cases 6a and 9b ($v_1^{\mathrm{in}}$, $v_2^{\mathrm{ex}}$, $v_{32}^{\mathrm{ex}}$ for case 6a and $v_1^{\mathrm{in}}$, $v_3^{\mathrm{in}}$, $v_1^{\mathrm{ex}}$ for case 9b). There exist two velocities in cases 9a and 10d ($v_1^{\mathrm{in}}$ and $v_2^{\mathrm{ex}}$ for 9a and $v_1^{\mathrm{ex}}$ and $v_{31}^{\mathrm{ex}}$ for 10d). For cases 1, 4, and 6b, we obtain the single velocity $v_1^{\mathrm{in}}$, whereas in cases 2a, 7a, 8b, and 10b, only a single velocity $v_1^{\mathrm{ex}}$ is found to exist. Finally, we note that no solutions for $v^{\mathrm{in}}$ or $v^{\mathrm{ex}}$ exist in cases 2b, 7b, 8a, 8c, 10a, and 10c.

%It should be noted that the boundary dividing regions of cases 9a and 9b, those of cases 2a and 2b, those of cases 8b and 8c, those of cases 10b and 10c, and those of cases 10b and 10d is also shown in Fig. \ref{FigTwo}. The boundary will be referred to as the boundary E, and the analytical form of this boundary will be obtained later. %Finally, the region where $C_2<0$ and $C_1>0$ has no $v^{\mathrm{in}}$ because the slip- and slip-velocity-hardening behaviors emerge there and the steady SFP is prohibited.

\begin{figure*}[tbp]
\centering
\includegraphics[width=16cm]{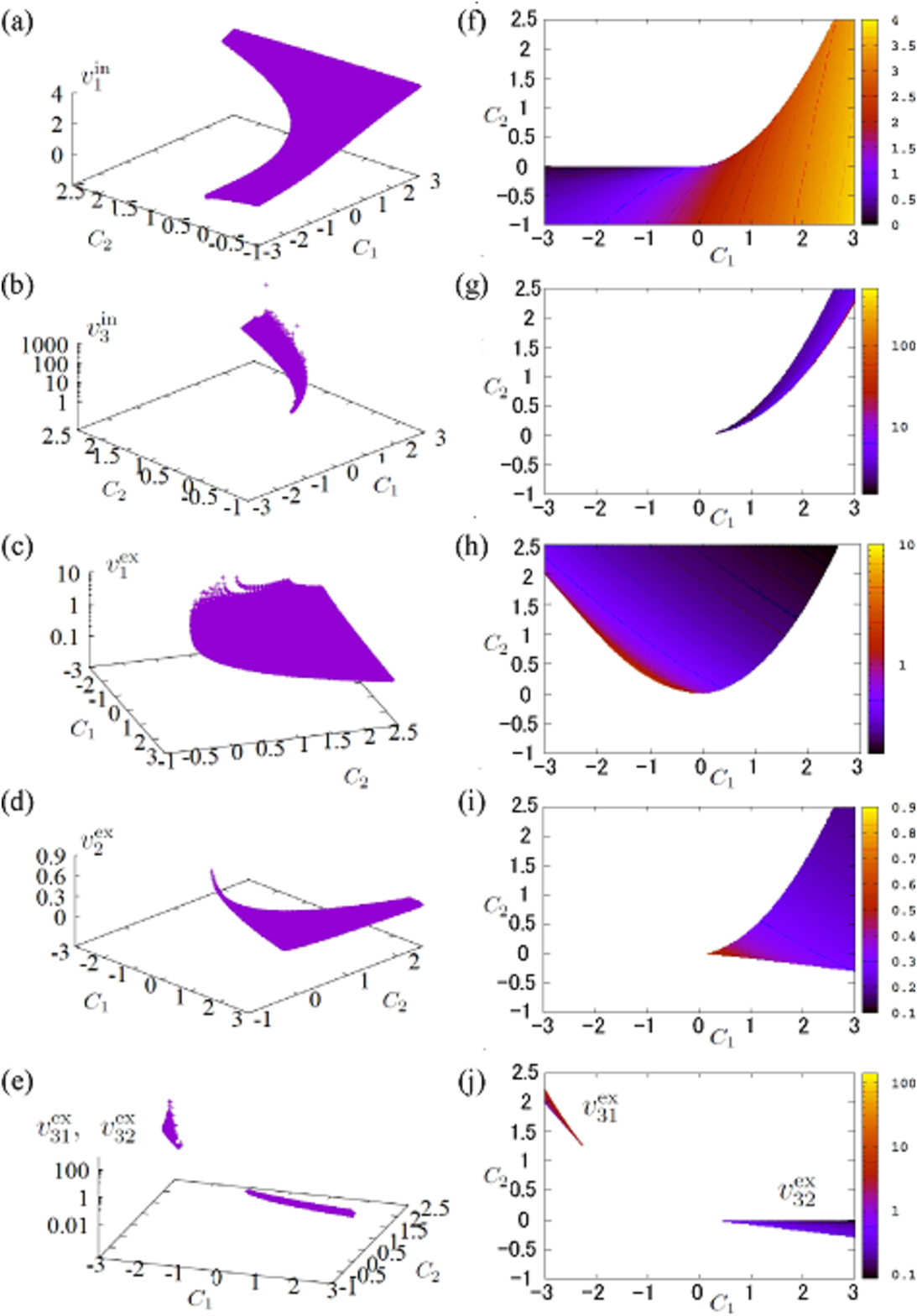}
\caption{The front velocities for various values of $C_1$ and $C_2$. (a) and (f) show the values of $v_1^{\mathrm{in}}$, (b) and (g) depict the values of $v_3^{\mathrm{in}}$, (c) and (h) show the values of $v_1^{\mathrm{ex}}$, (d) and (i) show the values of $v_2^{\mathrm{ex}}$, and (e) and (j) indicae tue values of $v_{31}^{\mathrm{ex}}$ and $v_{32}^{\mathrm{ex}}$. The 3D plots are shown in (a--e), while the contour maps are illustrated in (f--j). The log scale is used for the $z$ axes in (b), (c), and (e), and the color bars in (g), (h), and (j). Note that the direction of the axes differs between the subfigures (a--e), and the range of values associated with the color bar also differs between the subfigures (f--j).}
\label{FigVel}
\end{figure*}

The values of the front velocities $v_1^{\mathrm{in}}$, $v_3^{\mathrm{in}}$, $v_1^{\mathrm{ex}}$, $v_2^{\mathrm{ex}}$, $v_{31}^{\mathrm{ex}}$, and $v_{32}^{\mathrm{ex}}$ are shown in Fig.~$\ref{FigVel}$. It is found that $v_3^{\mathrm{in}}$, $v_1^{\mathrm{ex}}$, and $v_{31}^{\mathrm{ex}}$ diverge close to boundary E. This divergence is because the values of $k_3^{\mathrm{in}}$, $k_1^{\mathrm{ex}}$, and $k_3^{\mathrm{ex}}$ are equal to zero, while $\omega_3^{\mathrm{in}}$, $\omega_1^{\mathrm{ex}}$, and $\omega_3^{\mathrm{ex}}$ take non-zero values, on the boundary. We can therefore conclude that these velocities describe optical modes and represent unphysical propagations. In addition, we confirmed that the velocities $v_1^{\mathrm{in}}$ and $v_2^{\mathrm{ex}}$ take the values $\sqrt{1+C_1}+\sqrt{C_1}$ and $\sqrt{1+C_1}-\sqrt{C_1}$, respectively, on the $C_1$ axis for $C_1>0$. These values are consistent with those found in Ref.~\cite{Suz19} because the slip-dependence of the friction stress was negelected in that work, i.e., $C_2=0$. The value of $v_{32}^{\mathrm{ex}}$ is not consistent with $\sqrt{1+C_1}-\sqrt{C_1}$. We therefore conclude that $v_1^{\mathrm{in}}$ and $v_2^{\mathrm{ex}}$ represent the physical SFP velocities for the intruding and extruding fronts, respectively. We can also suggest that $v_1^{\mathrm{in}}$ and $v_2^{\mathrm{ex}}$ are the extension of the intruding and extrding front velocities obtained in Ref.~\cite{Suz19} to the model with the friction law depending on the slip and slip velocity.

%Moreover, we emphasize that $v_1^{\mathrm{in}}$ is the acoustic mode, while $v_3^{\mathrm{in}}$ corresponds to the optical mode, since $k_3^{\mathrm{in}}$ was found to be zero on the boundary between the cases 9a and 9b and hence $v_3^{\mathrm{in}}$ diverges at the boundary. Therefore, only $v_1^{\mathrm{in}}$ has physical meaning and can be observed for the intruding SFP. Actually, $v_1^{\mathrm{ex}}$ and $v_3^{\mathrm{ex}}$ are also the optical mode because $k_1^{\mathrm{ex}}=0$ on the boundary between the cases 2a and 2b, and $k_3^{\mathrm{ex}}=0$ on the boundary between the cases 6 and 9b and between the cases 10b and 10d. From Fig. \ref{FigVel}(e), (f), (i), and (j), we can observe that $v_1^{\mathrm{ex}}$ and $v_3^{\mathrm{ex}}$ diverge on these boundaries. We can conclude that only $v_2^{\mathrm{ex}}$ is reasonable for the extruding SFP. %; the velocity $v^{\mathrm{in}}_2$ does not exist because $(k_2^{\mathrm{in}})^2<0$. If the extruding front velocities exist, the negative solution for $\omega_i$ must exist. We can see that the relationships $v^{\mathrm{ex}} < v_s$ and $v_s < v^{\mathrm{in}}$ are not always satisfied. 

\section{NUMERICAL TREATMENT: IMPLICATIONS FOR SLIP BEHAVIOR} \label{secNT}

In this section we show the spatiotemporal evolutions of the slip velocity found via numerical computations using the values shown in Table \ref{Tcase}. We confirm that the values of $v_1^{\mathrm{in}}$ and $v_2^{\mathrm{ex}}$ obtained using the LMSH are accurate.

\begin{table}{}
\caption{Values of $C_1$ and $C_2$ as examples, and the resulting intruding and extruding front velocities for the various cases. The velocities in parentheses describe the optical modes.}
\begin{tabular}{cccl}%{100mm}
\hline\hline
Cases & $C_1$ & $C_2$ & Velocity \\ \hline
1 & $-0.375$ & $-0.03125$ & $v_1^{\mathrm{in}}$ \\
2a & 1 & 2 & $(v_1^{\mathrm{ex}})$ \\
2b & $-1$ & 0.2 &  \\
4 & 1 & $-0.5$ & $v_1^{\mathrm{in}}$ \\
6a & 1 & $-0.05$ & $v_1^{\mathrm{in}}, v_2^{\mathrm{ex}} \ (v_{32}^{\mathrm{ex}})$ \\
6b & $-2$ & $-0.05$ & $v_1^{\mathrm{in}}$ \\
7a & 1 & 0.5 & $(v_1^{\mathrm{ex}})$ \\
7b & $-2$ & 0.5 &  \\
8a & $-1$ & 0.05 &  \\
8b & $-3$ & 2.75 & $(v_1^{\mathrm{ex}})$ \\
8c & $-1.7$ & 0.71 &  \\
9a & 1 & 0.1 & $v_1^{\mathrm{in}}, v_2^{\mathrm{ex}}$ \\
9b & 1 & 0.3 & $v_1^{\mathrm{in}}, v_2^{\mathrm{ex}} \ (v_3^{\mathrm{in}})$ \\
10a & $-2$ & 0.1 &  \\
10b & $-3$ & 2.5 & $(v_1^{\mathrm{ex}})$ \\
10c & $-1.75$ & 0.755 &  \\
10d & $-3$ & 2.125 & $(v_1^{\mathrm{ex}}, v_{31}^{\mathrm{ex}})$ \\ \hline\hline
%\footnotetext[1]{Defined in the slip zone}
%\footnotetext[1]{Functions of time}
\end{tabular}
\label{Tcase}
\end{table}

\subsection{Intruding front propagation} \label{secIFP}

In this section we numerically consider the intruding front velocities. We consider $v_1^{\mathrm{in}}$ because $v_2^{\mathrm{in}}$ does not exist and $v_3^{\mathrm{in}}$ describes an optical mode, which represents an unphysical propagation. As such, we treat the cases 1, 4, 6a, 6b, 9a, and 9b (see Table \ref{Tcase}). We consider the EUL model, i.e., $F(u)=p_u-u$. In the following, we assume $p_u=1$.

For the numerical computations a friction law must be determined. There are few restrictions on the choice of friction laws provided they are reduced to the form of Eq.~(\ref{eqLin}) via linearization. Here, we adopt following expression for $\tau_{\mathrm{v}}(\dot{u})+\tau_{\mathrm{s}}(u)$:
%\begin{equation}
%\ddot{u}=u''+\dot{u}''+[-C_2 u +C_1 \dot{u}] [H(C_2 u -C_1 \dot{u})-H(C_2 u -C_1 \dot{u}-1)]. \label{eqvin}
%\end{equation}
%\begin{equation}
%\ddot{u}=u''+\dot{u}''-\tau_{\mathrm{s}}-\tau_{\mathrm{v}}, \label{eqvin}
%\end{equation}
\begin{eqnarray}
\tau_{\mathrm{v}}(\dot{u})+\tau_{\mathrm{s}}(u)=&1&-\frac{C_1}{2|C_1|} (1-\exp(-2|C_1| \dot{u})) \nonumber \\
&+& \frac{C_2-1}{2|C_2-1|} (1-\exp(-2|C_2-1|u)). \label{eqFSV}
\end{eqnarray}
%\begin{equation}
%\tau_{\mathrm{s}}=\frac{\sigma}{2} \left(1+\frac{2C_2}{|C_2|} (1-\exp(-|C_2|u)) \right) \label{eqITS}
%\end{equation}
%\begin{equation}
%\tau_{\mathrm{v}}=\frac{\sigma}{2} \left( 1-\frac{2C_1}{|C_1|} (1-\exp(-|C_1| \dot{u})) \right) \label{eqITV}
%\end{equation}
In this model, the system is in a critical state at the onset of end-loading, i.e., $F(0)-\tau_{\mathrm{v}}(0)-\tau_{\mathrm{s}}(0)=0$, as assumed in Sec.~\ref{secAT}. Additionally, it is possible to show that the right-hand side of Eq.~(\ref{eqFSV}) is always positive for any positive values of $u$ and $\dot{u}$, which is a required condition for the friction laws. We define the slip front as the region where $u$ and $\dot{u}$ are sufficiently small that only their linear terms exist in Eq.~(\ref{eqFSV}), and with this assumption, we can derive the relation %Additionally, note that with the limits $u \to \infty$ and $\dot{u} \to \infty$, we have $\tau_{\mathrm{v}}+\tau_{\mathrm{s}} \to 1$ if $(\mathrm{sign} (C_1), \mathrm{sign}(C_2))=(+,+)$ and $(-,-)$, $\tau_{\mathrm{v}}+\tau_{\mathrm{s}} \to 2$ if $(\mathrm{sign} (C_1), \mathrm{sign}(C_2))=(-,+)$, and $\tau_{\mathrm{v}}+\tau_{\mathrm{s}} \to 0$ if $(\mathrm{sign} (C_1), \mathrm{sign}(C_2))=(+,-)$ where $\mathrm{sgn}(\cdot)$ is the sign function. 
\begin{equation}
F(u)-\tau_{\mathrm{v}}(\dot{u})-\tau_{\mathrm{s}}(u) \sim C_1 \dot{u} -C_2 u
\end{equation}
from Eq.~(\ref{eqFSV}), which is seen to be equivalent to Eq.~(\ref{eqLin}). Though the friction law (\ref{eqFSV}) looks artificial, we emphasize that this expression is an example which permits numerical calculations and allows us to confirm the validity of the analytical treatment presented in this work.

%Additionally, we employ the stress-sign-change and the slip criterion. The maximum static friction stress required for the slip criterion is $-\sigma=-1$, as mentioned in Sec. \ref{secAT}.

To permit numerical calculations, the system must be finite, whereas an infinite system was assumed in the analytical treatment. We assume that the end-loading point is located at the position $x=-500$ for the finite system employed for the numerical calculations, and we use the value of $p_{-500} \equiv \partial u/\partial x|_{x=-500}=-0.05$ as the end-loading stress, instead of $p_e$ in the analytical work. In the numerical work, the Runge-Kutta method with fourth-order accuracy is used. %In actuality, we need not change this value because the governing equation (\ref{eqvin}) is linear and the value of $p_{-500}$ is irrelevant. However, note that the intruding front does not emerge for $C_2>0$ and $C_1<0$ (see Fig. \ref{FigTwo}). 

\begin{figure*}[tbp]
\centering
\includegraphics[width=12.5cm]{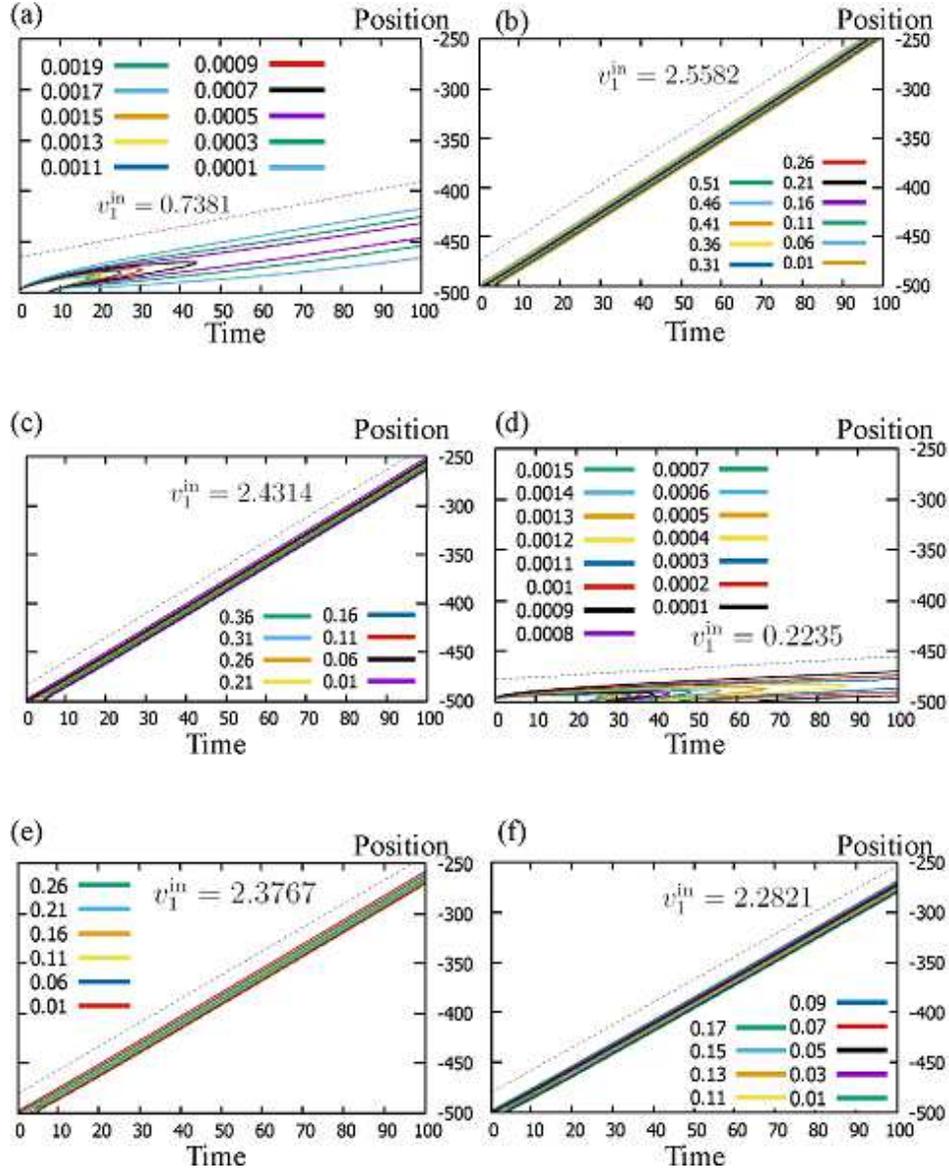}
\caption{The slip-velocity profiles obtained in this work. The color curves are isolines of the slip velocities, and the values with short color lines describe the values of the slip velocity on the isolines. The parameter values are (a) $C_1=-0.375$ and $C_2=-0.03125$ (case 1), (b) $C_1=1$ and $C_2=-0.5$ (case 4), (c) $C_1=1$ and $C_2=-0.05$ (case 6a), (d) $C_1=-2$ and $C_2=-0.05$ (case 6b), (e) $C_1=1$ and $C_2=0.1$ (case 9a), and (f) $C_1=1$ and $C_2=0.3$ (case 9b). The gradients of the dotted lines represent the analytically obtained values of the SFP velocities, which are described for each subfigure. } %The slip velocities in the yellow region in (e) and the blue region in (f) are almost zero. 
\label{FigSlip1}
\end{figure*}

\begin{figure}[tbp]
\centering
\includegraphics[width=8.5cm]{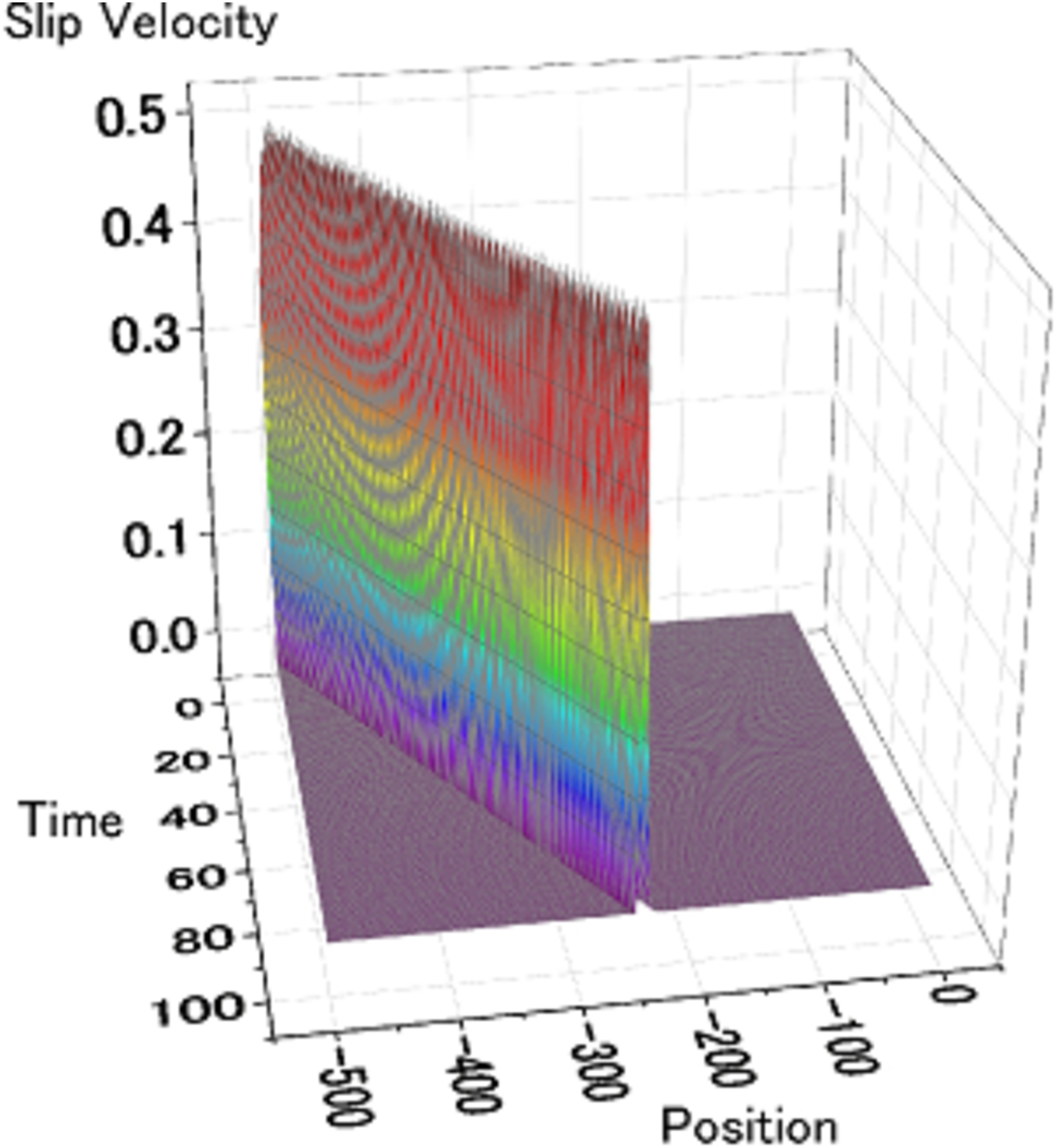}
\caption{The 3D plot for the spatiotemporal evolution with the case 4 [Fig.~\ref{FigSlip1}(b)].}
\label{FigVel3D}
\end{figure}

We can then numerically obtain the spatiotemporal evolutions of the slip velocities, which are shown in Fig.~$\ref{FigSlip1}$. Figure \ref{FigSlip1} shows that all the cases considered here generate intruding front velocities which are in agreement with those obtained analytically. This indicates that the LMSH is effective and that the normalized form of the governing equation accurately determines the SFP behavior. We note that the slip velocity profile is pulse-like in Fig.~\ref{FigSlip1}, which is clearly shown in Fig.~\ref{FigVel3D}. This result indicates that the slip velocity is always positive, and that the slip reversal does not occur when the stated friction law is considered [Eq.~(\ref{eqFSV})].

The supersonic front propagation is observed in Fig.~\ref{FigSlip1}(b), (c), (e) and (f); the shear wave velocity is unity in the present framework. These cases correspond to the slip- and/or slip-velocity-weakening behaviors with the critical condition just before the slip. Though the treatment in this study is not the same as that of the crack-tip propagation, the front propagation obtained here may be consistent with the supersonic crack-tip propagations addressed in previous studies \cite{Bue03, Bue06, Mar, Gia}. %the friction law (\ref{eqFSV}) is only an example, and 

%The faster one propagates with $v_1^{\mathrm{in}}$, whereas the SFP velocity for the other front is not $v_2^{\mathrm{ex}}$. This is because that the condition $u \ll 1$ is not satisfied at the slower front. %Generally, the extruding front does not emerge in these cases because if the intruding front exists, the extruding front is overtaken. %Moreover, the consistency of the numerical results with the analytical ones is important for the nonlinear wave propagation.% It is also important that the slip at the loading point diverges in the cases 1, 4, 6a, 6b, and 9a, indicating macroscopic slip generation.

%For the cases shown in Fig. $\ref{FigSlip1}$(b)-(d), the real solutions for $v^{\mathrm{in}}$ exist, while in Fig. $\ref{FigSlip1}$(e)-(g) we have no real solutions for $v^{\mathrm{in}}$. Additionally, the case $\Gamma$ has two solutions $v_1^{\mathrm{in}}$ and $v_3^{\mathrm{in}}$, and the cases A and B have only the single solution $v_1^{\mathrm{in}}$. We regard $x=-500$ as the loading point and the value $p_{-500} \equiv p(-500)$ is fixed to be $-10^{-3}$ as the boundary condition for all cases. We first consider the cases A and B (Figs. $\ref{FigSlip1}$(b) and (c)). If we have the single solution for $v^{\mathrm{in}}$, the steady front propagation can be observed, and the front propagates with the velocity $v_1^{\mathrm{in}}$ as confirmed by the dark straight line. It is also important that the slip at the loading point diverges in these cases, indicating macroscopic slip generation.

\subsection{Extruding front propagation}  \label{secEFP}

Here we consider the extruding front velocities. We consider only $v_2^{\mathrm{ex}}$ because $v_1^{\mathrm{ex}}$ and $v_{31}^{\mathrm{ex}}$ correspond to the optical modes, and the value of $v_{32}^{\mathrm{ex}}$ is not consistent with that of previous study. To simulate the extruding front, we consider the EL model \cite{Suz19}. %In \cite{Suz19}, the extruding front propagation was simulated with the friction law quadratically depending on the slip velocity; i.e., $C_2=0$. The dependence on the slip velocity is extended here to the slip-dependence.

In this numerical work, we assume the following friction laws,
%\begin{eqnarray}
%\tau_{\mathrm{v}}(\dot{u})+\tau_{\mathrm{s}}(u)=&C_v& \dot{u} (C'_1-\dot{u}) [H(\dot{u})-H(\dot{u}-C'_1)] \nonumber \\
%&+&C_s u (C'_2-u) [H(u)-H(u-C'_2)], \label{eqETSVQ}
%\end{eqnarray}
\begin{equation}
\tau_{\mathrm{v}}(\dot{u})=C_v \dot{u} (C'_1-\dot{u}) [H(\dot{u})-H(\dot{u}-C'_1)] \label{eqETVQ}
\end{equation}
and
\begin{equation}
\tau_{\mathrm{s}}(u)=C_s u (C'_2-u) [H(u)-H(u-C'_2)], \label{eqETSQ}
\end{equation}
where $C_v, \ C_s, \ C'_1$, and $C'_2$ are positive constants. The friction laws given in Eqs.~(\ref{eqETVQ}) and (\ref{eqETSQ}) assume a quadratic dependence of the friction stress on the slip velocity and slip, respectively. The quadratic dependence of the slip velocity was assumed in Ref.~\cite{Suz19}, and this dependence is now extended to the slip dependence. In the following discussion it is shown that the friction laws in Eqs.~(\ref{eqETVQ}) and (\ref{eqETSQ}) generate the extruding front. %We consider the combination of Eqs. (\ref{eqETSQ}) and (\ref{eqETVQ}), $\tau_{\mathrm{sv}} \equiv \tau_{\mathrm{s}} +\tau_{\mathrm{v}}$.  and $C_A \equiv C_s C'_2$ and $C_B \equiv C_v C'_1$ are assumed to be constant

First, we demonstrate that the friction laws (\ref{eqETVQ}) and (\ref{eqETSQ}) do not generate the intruding front propagation. These friction laws state that the system corresponds to $C_1<0$ and $C_2>0$ when $u \ll 1$ and $\dot{u} \ll 1$. As shown in Fig.~\ref{FigTwo}(a), $v_1^{\mathrm{in}}$ does not exist in the region in which $C_1<0$ and $C_2>0$. Therefore, under these conditions we do not have an intruding front propagation.  %One is that the laws can describe the non-critical state. 

Here we demonstrate how the extruding front velocity is determined with the friction laws given in Eqs.~(\ref{eqETVQ}) and (\ref{eqETSQ}). We first introduce the variable,
\begin{equation}
\dot{w} \equiv C''_1 -\dot{u}, \label{eqw2d}
\end{equation}
where $C''_1$ is a positive constant. From the definition in Eq.~(\ref{eqw2d}), we obtain,
\begin{equation}
w =C''_1 t+W(x) -u,
\end{equation}
where $W(x)$ is an arbitrary function of $x$. We consider the region near the slip front propagating with a constant velocity. Therefore, if we assume that $C''_1$ is equal to the extruding front velocity, we can see that $C''_1 t+W(x)$ is a constant, which results in 
\begin{equation}
w = C''_2 -u, \label{eqw2}
\end{equation}
where $C''_2$ is a positive constant. We assume that $w$ and $\dot{w}$ are sufficiently small to mean that the terms of order $w^2$ and $\dot{w}^2$ and higher can be neglected. Using Eqs.~(\ref{eqNUFG}), (\ref{eqETVQ}), (\ref{eqETSQ}), (\ref{eqw2d}), and (\ref{eqw2}), considering only the linear terms of $w$ and $\dot{w}$, we obtain the governing equation:
\begin{eqnarray}
\ddot{w}=w''&+&\dot{w}'' \nonumber \\
&+&C_v (C'_1-2C''_1) \dot{w} [H(C''_1-C'_1-\dot{w})-H(C''_1-\dot{w})] \nonumber \\
&+&C_s (C'_2-2C''_2)w [H(C''_2-C'_2-w)-H(C''_2-w)]. \nonumber \\
\end{eqnarray}
From these expressions, we conclude that the values $C_1$ and $C_2$ in Eq.~(\ref{eqeom4}) are given by:
\begin{equation}
C_1=C_v (C'_1-2C''_1),
\end{equation}
and
\begin{equation}
C_2=-C_s (C'_2-2C''_2).
\end{equation}

Here, we demonstrate a method determining the values of $C''_1$ and $C''_2$. First, we write $v_2^{\mathrm{ex}}$ as a function of $C_1$ and $C_2$, i.e., $v_2^{\mathrm{ex}}(C_1, C_2)$. This function is illustrated in Fig.~\ref{FigVel}(i). We then define the values of $C'''_1 \equiv -\partial \tau_{\mathrm{v}} (\dot{u})/\partial \dot{u}|_{\dot{u}=C''_1}$ and $C'''_2 \equiv \partial \tau_{\mathrm{s}} (u)/\partial u|_{u=C''_2}$. Since the quantities $-C'''_1$ and $C'''_2$ are the gradients of $\tau_{\mathrm{v}}(\dot{u})$ and $\tau_{\mathrm{s}}(u)$, respectively, at $\dot{u}=C''_1$ and $u=C''_2$, we obtain
\begin{equation}
\tau_{\mathrm{v}}(\dot{u})=-C'''_1 (\dot{u}-C''_1)=C'''_1 \dot{w}
\end{equation}
and
\begin{equation}
\tau_{\mathrm{s}}(u)=-C'''_2 (u-C''_2) =-C'''_2 w,
\end{equation}
around $\dot{u}=C''_1$ and $u=C''_2$. Therefore, the equation
\begin{equation}
v_2^{\mathrm{ex}}(C'''_1, C'''_2)=C''_1 \label{eqVex1}
\end{equation}
yields a relation between $C''_1$ and $C''_2$. Since $C'''_1$ depends on $C''_1$, the above expression represents a self-consistent equation. However, we cannot determine the two unknown quantities, $C''_1$ and $C''_2$, uniquely from this single equation. We obtain a second expression relating the two parameters via consideration of $p_e$. We assume that the single relation
\begin{equation}
G(C'''_1, C'''_2, p_e)=0 \label{eqVex2}
\end{equation}
exists, whereas obtaining the analytic form of $G$ is impossible. However, we can expect that a larger value of $|p_e|$ induces a larger value of $v_2^{\mathrm{ex}}$. Equations (\ref{eqVex1}) and (\ref{eqVex2}) determine the extruding front velocity $v_2^{\mathrm{ex}}(=C''_1)$ for a given $p_e$. Finally, if these equations do not have the solution, the steady SFP does not exist. %The parameter $p_e$ also plays an important role to determine the extruding front velocity because the system is nonlinear. The extruding front velocity can be self-consistently determined, while analytical treatment is impossible. 

Here, we show how the analytical treatment and numerical calculations are consistent with Eqs.~(\ref{eqETVQ}) and (\ref{eqETSQ}). First, we set the values of $C_v, \ C_s, \ C'_1, \ C'_2$, and $p_{-500}$. We then numerically compute the spatiotemporal evolutions of the slip, and the slip front is roughly determined from the slip-velocity profile. We also roughly evaluate the slip and slip velocity at the front; these values correspond to $C''_2$ and $C''_1$, respectively. With $C''_1$, $C''_2$ obtained, using Eqs.~(\ref{eqETVQ}) and (\ref{eqETSQ}) we obtain the values of $C'''_1$ and $C'''_2$; thus we can obtain $v_2^{\mathrm{ex}}$ because it is a function of $C'''_1$ and $C'''_2$. We then estimate the SFP velocity at the determined front from the numerical computations; the velocity obtained using this estimate is denoted $v_\mathrm{est}$. We finally confirm that $v_\mathrm{est}$ is approximately equal to $v_2^{\mathrm{ex}}(C'''_1, C'''_2)$. %The obtained SFP velocity corresponds to $C''_1$. In addition, the slip at the front is roughly estimated, and the value is considered to be $C''_2$. 

It remains, however, to confirm another condition: as observed, in the analytical treatment, the inequalities $0 \le \dot{u} \le C'_1$ and $0 \le u \le C'_2$ must be satisfied at the slip front. These conditions are satisfied when $0 \le C''_1 \le C'_1$ and $0 \le C''_2 \le C'_2$, in the numerical treatment. It should be confirmed that these relations are satisfied in the numerical calculations. %Moreover, we also adopt the stress-sign-change and the slip criterion. However, note that the maximum static friction stress is zero in the present model because $\lim_{\dot{u} \to 0} \tau_{\mathrm{v}}(\dot{u})=\lim_{u \to 0} \tau_{\mathrm{s}}(u)=0$. Therefore, we consider only the stress-sign-change.

\begin{figure*}[tbp]
\centering
\includegraphics[width=13cm]{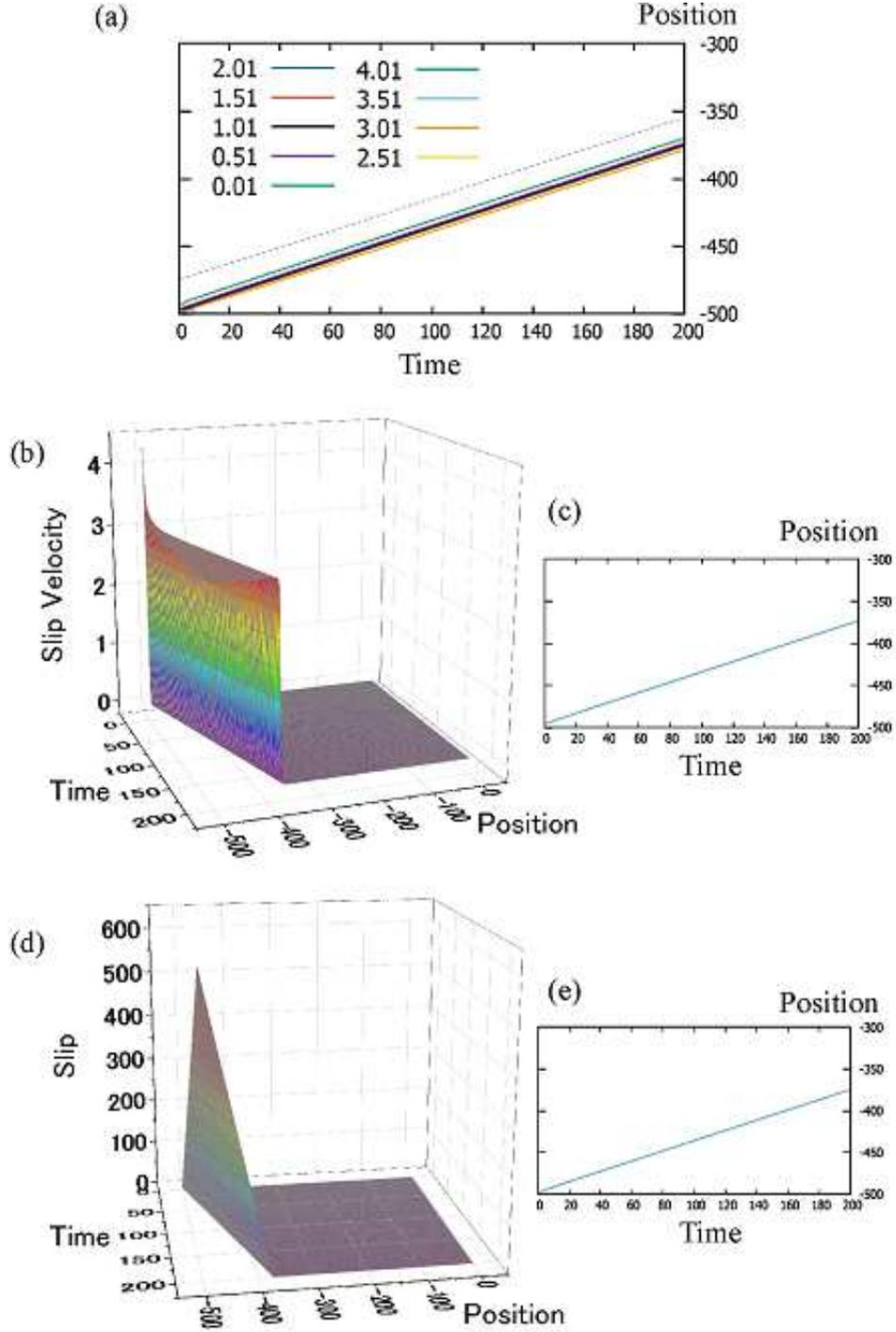}
\caption{Spatiotemporal evolution of the slip and slip velocity in the case of $C_v=1, \ C_s=0.2, \ C'_1=1, \ C'_2=5$, and $p_{-500}=-5$. (a) A contour map showing lines of constant slip velocity. The estimated SFP velocity (the gradient of the dotted straight line) is taken to be 0.604. (b) 3D plot for the spatiotemporal evolution of the slip velocity. (c) Plot for the point with $\dot{u}=0.3$. (d) 3D plot for the spatiotemporal evolution of the slip. (e) Plot for the point with $u=2.5$.}
\label{FigSlip2}
\end{figure*}

We consider an example with $C_v=1, \ C_s=0.2, \ C'_1=1, \ C'_2=5$, and $p_{-500}=-5$, and we apply the procedure described above (Fig.~\ref{FigSlip2}). In this case, as shown in Fig.~\ref{FigSlip2}(b), the gradient of the slip velocity profile along the $x$ axis is so large that it is not possible to determine the precise values of $C''_1$ and $C''_2$. Therefore, we can only use an estimate for their values based on the profiles of the slip and slip velocity; in this case, we take $C''_1=0.3$ and $C''_2=2.5$ [see Fig.~\ref{FigSlip2}(b-e)]. The SFP velocity is relatively insensitive to variations around the values of $C''_1$ and $C''_2$ adopted here. These $C''_1$ and $C''_2$ values together with Eqs.~(\ref{eqETVQ}) and (\ref{eqETSQ}) give the values $C'''_1=0.4$ and $C'''_2=0$, which lead to $v_2^{\mathrm{ex}}(0.4, 0)=0.55$. The SFP velocity obtained from the gradient of the dotted line in Fig.~\ref{FigSlip2}(a) is estimated to be $v_{\mathrm{est}}=0.604$. This almost coincides with the value of $v_2^{\mathrm{ex}}(0.4, 0)=0.55$ obtained above. Additionally, the conditions $0 \le C''_1 \le C'_1$ and $0 \le C''_2 \le C'_2$ are satisfied in this calculation. Thus, we have found good agreement between the analytic and numerical approaches in the obtained values of the extruding front velocity. Finally, we note the slip profile shown in Fig.~\ref{FigSlip2}(b) leads to the conclusion that the slip reversal does not occur in this case. %This almost coincides with $v_{\mathrm{est}}=0.604$, though the difference between the analytical result and numerical one emerges because of the finiteness of the system for the numerical calculations, which emerged also in Ref. \cite{Suz19}. This almost coincides with $v_{\mathrm{est}}=0.604$, though the difference between the analytical result and numerical one emerges because of the finiteness of the system for the numerical calculations, which emerged also in Ref. \cite{Suz19}. %Actually, even if we fix $C_s, C_v, C'_2$ and $C'_1$, we cannot determine $C''_2$ and $C''_1$ a priori. Let us consider the example case $C_s=1, \ C'_2=5, \ C_v=1$, and $C'_1=1$. With these values, the SFP velocity is determined by the following conditions. The values of $\partial \tau_{\mathrm{s}}(u)/\partial u|_{u=C'_2}$ and $-\partial \tau_{\mathrm{v}} (\dot{u})/\partial \dot{u}|_{\dot{u}=C'_1}$ are $-2$ and $-1$, respectively. However, this case does not correspond to the case $C_2=2$ and $C_1=-1$ (case 2b). case 6a

\section{DISCUSSION AND CONCLUSIONS} \label{secDisCon}

In this work we have considered an infinite 1D viscoelastic block on a substrate subject to end-loading stress and the end- and upper-loading stresses together with a friction law that depends on the slip and slip velocity. The two intruding front velocities and three extruding front velocities were analytically obtained based on the LMSH, which requires only the linearized governing equation. Of these five velocities, three represent the optical modes, and the unique intruding and extruding front velocities were found to exist for physically meaningful SFPs. Via numerical calculations, the intruding front velocity was obtained for the EUL model, and the extruding front velocity was realized for the EL model; these results indicated that the detail of the dependence of the friction stress on the slip and the slip velocity does not affect the front velocity, and that the linearized governing equation is sufficient to obtain accurate results.

Our framework is found to be quantitatively consistent with the results of previous studies. For example, the dependences of $v_1^{\mathrm{in}}$  and $v_2^{\mathrm{ex}}$ on $C_1$ for $C_2=0$ are consistent with those of $v^{\mathrm{in}}$ and $v^{\mathrm{ex}}$ in Ref. \cite{Suz19}, respectively, as mentioned in Sec.~\ref{secAT}. Additionally, the form of $v_1^{\mathrm{in}} (C_1)$ for $C_2=1$ is consistent with that obtained in Myers and Langer \cite{Mye}. Though they obtained only an approximate solution, the exact analytical solution has been obtained here. Furthermore, note that their model corresponds to the case $C_2=1$, and they changed parameters $\eta$ and $\alpha$, where $\eta$ is the viscosity, and $\alpha$ corresponds to $C_1$. Therefore, both their and our models have two parameters. However, our model is superior because we can treat both positive and negative $C_2$ in a single framework. Moreover, we were successful in explaining the variation in SFP velocities using only friction laws with constant viscosity. %Additionally, we obtain $C_1=58.8$ and $C_2=14.9$ using parameter values employed in the model A of Otsuki and Matsukawa \cite{Ots}. Using these values, $v_1^{\mathrm{in}}$ is calculated to be approximately $25$, which roughly equals to the propagation velocity of the bounded rapid precursors obtained in their numerical calculation. There exists a difference, however, between the model considered in Ref. \cite{Ots} and that considered here in that the normal stress is not a constant value in Ref. \cite{Ots}. Therefore, the treatment in the present paper cannot be applied directly to their study, but we can roughly evaluate the front propagation velocity using the framework provided here. %Finally, our framework is an extension of the works by Tr\o mborg $et \ al.$ \cite{Tro} and Amundsen $et \ al.$ \cite{Amu2} because they did not consider the slip-dependent and the slip-velocity-dependent friction laws. %Moreover, our framework has advantage to previous studies such as \cite{Amp, Gab} because our result is based on the numerical treatment. %(1) the criticality is not realized in their model, 

We also have shown the regions in which the SFP velocities exist in the $C_1-C_2$ phase space; we also have presented the analytical forms of the boundaries of these regions in the $C_1-C_2$ phase space. This phase space may be useful to predict, e.g., whether ordinary or slow earthquakes are likely in natural faults, if the friction law is determined.

Since LMSH linearizes the governing equations, it cannot give an estimation of the macroscopic slip profile. Note that the pulse-like slip was observed in Fig.~\ref{FigVel3D}, and the step-function-like slip emerged in Fig.~\ref{FigSlip2}. These are consistent with previous laboratory experiments \cite{Nie} or seismological observations \cite{Das, Aag}. Nonetheless, predicting these slip behaviors is difficult with the present framework.

Finally, the critical state was assumed for the EUL model. Based on this assumption, we can proceed with the analytical treatment. In reality, the SFP in a non-critical state has been analyzed \cite{Mye, Bar12}, and the SFP velocities have been obtained. They are, however, approximations. Though the current treatment cannot be directly applied to the non-critical case, we anticipate that it will be an important step in analytically treating SFP velocities with the non-critical state.

%Our model can be applied to an isotropic and homogeneous system, so that precursors are expected to be explained in terms of our model.

\appendix

\renewcommand{\theequation}{A\arabic{equation}}
\setcounter{equation}{0}

\section{Exact Solutions of $f_{\mathrm{in}}(\omega_i)=0$} \label{secAA}

Here, we present details of the analytical treatment of $f_{\mathrm{in}}(\omega_i)=0$. This is the cubic equation, and we can solve it analytically. We first define the variables $D_1$ and $D_2$:
\begin{equation}
D_1 = -\left( \frac{16 C_1^3}{27} +\frac{2C_1}{3} (C_1-3C_2)-2C_2 \right) \label{eqAC1}
\end{equation}
and
\begin{equation}
D_2=-\frac{1}{27} \left( -\frac{4 C_1^2}{3} -C_1+3C_2 \right)^3.  \label{eqAC2}
\end{equation}
Using Eqs.~(\ref{eqAC1}) and (\ref{eqAC2}), we define
\begin{equation}
S_{\pm} = \frac{-D_1 \pm \sqrt{D_1^2-4D_2}}{2}
\end{equation}
(double signs in the same order), which can take complex values. With these values and the cubic root of unity, $\omega_0=(-1+ i \sqrt{3})/2$, we obtain:
\begin{equation}
\omega^{\mathrm{in}}_1=S_+^{\frac{1}{3}}+S_-^{\frac{1}{3}} +\frac{2C_1}{3}, \label{eqomegain1}
\end{equation}
\begin{equation}
\omega^{\mathrm{in}}_2=\omega_0 S_+^{\frac{1}{3}} + \omega_0^2 S_-^{\frac{1}{3}} +\frac{2C_1}{3}, \label{eqomegain2}
\end{equation}
\begin{equation}
\omega^{\mathrm{in}}_3=\omega_0^2 S_+^{\frac{1}{3}} + \omega_0 S_-^{\frac{1}{3}} +\frac{2C_1}{3}. \label{eqomegain3}
\end{equation}
If $S_{\pm}$ is real and negative, $S_{\pm}^{1/3}$ is defined as $-|S_{\pm}|^{1/3}$. Therefore, $\omega^{\mathrm{in}}_1$ always gives a real-valued solution. With Eqs.~(\ref{eqomegain1})$-$(\ref{eqomegain3}) and Eq.~($\ref{eqk2}$), we can define $k_j^{\mathrm{in}} \equiv ((-2C_2+C_1 \omega_j^{\mathrm{in}})/\omega_j^{\mathrm{in}})^{1/2}$ and $v_j^{\mathrm{in}} \equiv \omega_j^{\mathrm{in}}/k_j^{\mathrm{in}}$ ($j=1,2,3$). We note that if $({k_j^{\mathrm{in}}})^2<0$, we can conclude that $v_j^{\mathrm{in}}$ does not exist. We can use exactly the same method to solve $f_{\mathrm{ex}} (\omega_i)=0$ and to define $\omega_j^{\mathrm{ex}}, \ k_j^{\mathrm{ex}}$ and $v_j^{\mathrm{ex}}$.

%\renewcommand{\theequation}{B\arabic{equation}}
%\setcounter{equation}{0}
%\section{Slip Velocity Profiles} \label{secAB}

\acknowledgments
T. S. was supported by JSPS KAKENHI Grant Number JP20K03771, and JSPS KAKENHI Grant Number JP16H06478 in Scientific Research on Innovative Areas ``Science of Slow Earthquakes.'' This study was supported by the Ministry of Education, Culture, Sports, Science and Technology (MEXT) of Japan, under its The Second Earthquake and Volcano Hazards Observation and Research Program (Earthquake and Volcano Hazard Reduction Research). This study was supported by ERI JURP 2020-G-08, 2021-G-02 and 2022-G-02 in Earthquake Research Institute, the University of Tokyo. %The author would like to thank Enago (www.enago.jp) for the English language review.

\end{document}